\documentclass[nofootinbib]{revtex4-2}

\usepackage{scrextend}
\usepackage{graphicx}
\usepackage{indentfirst}
\usepackage{latexsym} 
\usepackage{multirow}
\usepackage{epsfig}
\usepackage{hyperref}
\usepackage{color}
\usepackage[usenames,dvipsnames]{xcolor}
\usepackage{amssymb}
\usepackage{amsfonts}
\usepackage{amsmath}
\usepackage{bm}
\usepackage{mathrsfs}
\usepackage{cleveref}
\usepackage{algorithm}
\usepackage{algorithmic}
\usepackage{subcaption}

\algsetup{indent=2em}

\usepackage[normalem]{ulem}
\useunder{\uline}{\ul}{}

\newcounter{comment}

\newcommand{\algword}[1]{{\ \mathbf{#1}\ }}

\makeatletter
\newcommand{\multiline}[1]{%
  \begin{tabularx}{\dimexpr\linewidth-\ALG@thistlm}[t]{@{}X@{}}
    #1
  \end{tabularx}
}
\newcommand{\bea}{\begin{eqnarray}}
\newcommand{\eea}{\end{eqnarray}}
\makeatother

\newcommand{\be}{\begin{equation}}
\newcommand{\ee}{\end{equation}}

\def\addFaMAF{Facultad de Matem\'atica, Astronom\'ia, F\'isica y Computaci\'on,\\
 Universidad Nacional de C\'ordoba, (5000) C\'ordoba, Argentina}

\def\addCONICET{CONICET}

\def\addIsistan{ISISTAN-CONICET Research Institute, \\UNICEN University, (7000) Tandil, Buenos Aires, Argentina}

\begin{document}

\title{An automated parameter domain decomposition approach for gravitational wave surrogates using hp-greedy refinement}

\author{Franco Cerino}
\affiliation{\addFaMAF}
\affiliation{\addCONICET}

\author{J. Andrés Diaz-Pace}
\affiliation{\addIsistan}

\author{Manuel Tiglio}
\affiliation{\addFaMAF}
\affiliation{\addCONICET}

\begin{abstract}
We introduce hp-greedy, a refinement approach for building gravitational wave surrogates as an extension of the standard reduced basis framework. Our proposal is data-driven, with a domain decomposition of the parameter space, local reduced basis, and a binary tree as the resulting structure, which are obtained in an automated way. When compared to the standard global reduced basis approach, the numerical simulations of our proposal show three salient features: 
i) representations of lower dimension with no loss of accuracy,
ii) a significantly higher accuracy for a fixed maximum dimensionality of the basis, in some cases by orders of magnitude, and 
iii) results that depend on the reduced basis seed choice used by the refinement algorithm.
We first illustrate the key parts of our approach with a toy model and then present a more realistic use case of gravitational waves emitted by the collision of two spinning, non-precessing black holes.  We discuss performance aspects of hp-greedy, such as overfitting with respect to the depth of the tree structure, and other hyperparameter dependences.
As two direct applications of the proposed hp-greedy refinement, we envision:
i) a further acceleration of statistical inference, which might be complementary to focused reduced-order quadratures, and
ii) the search of gravitational waves through clustering and nearest neighbors.
\end{abstract}

\maketitle

\section{Introduction} \label{sec:intro}

Gravitational waves (GWs) are perturbations of space-time produced by massive accelerating objects, predicted by Einstein's equations for General Relativity (GR). Since 2015, when they were directly measured for the first time~\cite{Abbott_2016}, they have become a new window to the universe, in addition to electromagnetic radiation.

When studying GWs, the ones produced by coalescing compact binary systems are widely considered. These are the strongest ones and hence the easiest to measure. Also, they provide information of the strong field regime of gravity. To model the waves emitted when the merger in a binary coalescence occurs, Numerical Relativity (NR) is needed, a task that requires expensive computational power. For example, a single simulation of a binary black hole system can cost $10^{4}-10^{5}$ CPU hours~\cite{Lehner_2014}. Furthermore, some studies are only feasible if the GWs can be computed fast enough, in real or quasi-real time, as in parameter estimation~\cite{Veitch_2015,Christensen_2022, Canizares:2013ywa} and matched filtering~\cite{Owen:1998dk,Cannon_2012,Babak_2013}. Therefore, developments that enable model evaluation and analysis in a fast and accurate way are necessary.

The use of the reduced basis method~\cite{Jan-RB,Chen:2010:CRB:1958598.1958625,Field:2011mf,prud'homme:70,quarteroni2015reduced} has been largely adopted in GW science~\cite{Varma:2019csw,gwsurr,catalog,PhysRevD.95.104023,PhysRevLett.115.121102,PhysRevD.96.024058,PhysRevX.4.031006}, both for building surrogate models and for statistical inference, significantly reducing the associated computational costs while retaining high accuracy --see~\cite{tiglio2021reduced} for a review. Existing approaches in this application domain have normally used a global basis, similar in spirit to spectral methods. In fact, the reduced basis approach is sometimes referred to as a {\it domain-specific spectral expansion}. However, there are cases in which being able to partially localize the basis is useful and might have significant performance effects. In this paper we do so by proposing an hp-greedy refinement approach~\cite{Eftang:2010}. From a numerical relativity perspective, hp-greedy is similar to spectral elements~\cite{Sarbach2012} but partitioning the parameter domain instead of the physical one (space-time). As we discuss, this strategy has several advantages when compared to a global approach, most prominently: a higher accuracy basis for the same number of elements, faster surrogate evaluations, and faster statistical inference. 

The organization of this paper is as follows. In Section~\ref{sec:hp} we describe in detail the hp-greedy reduced basis framework, its algorithms and supporting notation. In Section~\ref{sec:toy} we illustrate its application to a toy model with an intended strong discontinuity in the parameter space. In Section~\ref{sec:gws} we apply the framework to the case study of the gravitational waves emitted by the collision of two spinning, non-precessing black holes in an initial quasi-circular orbit. We close in Section~\ref{sec:com} with comments and possible future directions of research. 

\section{hp-greedy reduced basis} \label{sec:hp}

The hp-greedy approach leverages the standard reduced basis method, building a partitioning (h-refinement) of the parameter space and a reduced basis (p-refinement, here denoted as RB) for each partition.

A global reduced basis is initially built. It this basis is not as accurate and compact as wanted, the domain is partitioned and new local bases are built in each subdomain. The idea is to find reduced bases for spaces with lower complexity by means of a divide-and-conquer strategy.

The partitioning is adaptive and recursive. It depends on how the structure of the solutions vary in parameter space, allowing to focus the partitioning where it is needed, stopping when the local basis is sufficiently accurate or after a maximum number of partitions. 

This section explains the h and p refinement procedures (subsections \ref{sec:p} and \ref{sec:h}) and how they work in synergy to obtain the hp-greedy approximation (subsection \ref{sec:alg}).

\subsection{p-refinement} \label{sec:p}

This is the standard reduced basis method (RBM). It is referred to as {\it p-refinement} borrowing language from spectral methods \cite{Sarbach2012,hesthaven2007spectral}, where the basis are polynomials (thus, the ``p'') and the {\it refinement} part refers to the property that the representation error decreases as the degree of the polynomial increases. By analogy, then {\it p-refinement} in this context means that the error of the reduced basis representation decreases as the dimensionality $n$ of the basis increases. 

The RBM is traditionally targeted to computationally-intensive parametrized problems which require multiple queries. It is an alternative to repeatedly solving the full problem, which might not be feasible or realistic in practice, allowing for compact and accurate representations of the elements under study.  In the case of GW modeling, the golden standard for solving the full problem is numerical solutions of the Einstein equations on supercomputers, which tends to be a remarkably difficult and expensive task. The RBM has allowed the construction of surrogate predictive models which are essentially indistinguishable from numerical relativity supercomputer simulations but can be evaluated in less than a second on a standard laptop. 

The RBM starts with a solution space of functions ${\cal F}:= \{ h_{\lambda}:=h_{\lambda}(t) := h(\lambda, t) \}$, where 
$\lambda$ is a parameter, in general multi-dimensional, in a compact domain $D$. In our application, $h(t)$ is a complex time series: a gravitational wave. More specifically, $\lambda$ can denote, for example, the masses and spins of each black hole in a binary collision. 

A sampling of ${\cal F}$ is used to form a training set ${\cal T}:= \{h_i = h_{\lambda_i}, \, i=1\ldots m \}$ for a number $m$ of parameter values $\{ \lambda_i\}_{i=1}^m$. The training set is used to build a compact reduced basis $\{e_i \}_{i=1}^n$, with $n << m$ in general, the compression ratio being $n/m$, which represents $\cal T$ through its linear span of the form
$$
h(t) \approx \sum_{i=1}^n c_i e_i (t) \, , 
$$
and furthermore, through validation, the original space $\cal F$ within a prescribed accuracy $\epsilon$ or a maximum dimension $n_{max}$. The choice of the coefficients $\{c_i \}$ should be such that the approximation is optimal, o quasi-optimal, in a precise mathematical sense. 

\subsubsection{Searching for an optimal reduced basis}
A basis of dimension $n$, being optimal with respect to the maximum error in the parameter space, is characterized by the Kolmogorov $n$-width \cite{Pinkus},
\begin{equation}
\label{eqn:kolmogorov_measure}
d_{n} := \min_{ \{e_{i} \}_{i=1}^n}  \max_{\lambda \in D} \ \min _{c_{i,\lambda} \in 	\mathbb{C}}  \ \|   h_{\lambda}(\cdot) - \sum\limits_{i=1}^n c_{i,\lambda} e_{i}(\cdot) \| ^2\, .
\end{equation}
It measures the maximum representation error in parameter space, given by an optimal basis and the optimal coefficients $\{ c_{i,\lambda} \}$, in the norm $ \|\cdot \| $. In our case, the latter is given by
\begin{equation}
\label{eqn:l2norm}
 \| h_{\lambda}(\cdot) \| ^2 := \int_{t_i}^{t_f} dt \   | h_{\lambda}(t) |^2, 
\end{equation}
inherited from the scalar product
\be
\langle h_1, h_2 \rangle:=  \int_{t_i}^{t_f} \bar{h}_i(t) h_f(t) \,  dt \, ,   \label{eq:dot}
\ee
where the bar indicates complex conjugation. 

If a basis is fixed in Equation (\ref{eqn:kolmogorov_measure}), the first minimization problem, i.e., the minimum over the coefficients $\{ c_{i, \lambda} \}$, turns to be a least squares one for those coefficients. There is a unique solution to this problem: it is the orthogonal projection $\mathcal{P}_n$ with respect to the scalar product (\ref{eq:dot}) to the span of the reduced basis \cite{tiglio2021reduced}. The n-width then takes the form
\begin{equation}
d_{n} = \min_{ \{e_{i} \}_{i=1}^n}  \max_{\lambda \in D}   \ \|   h_{\lambda}(\cdot) - \mathcal{P}_n h_{\lambda}(\cdot)  \| ^2 \, .
\end{equation}

In some cases, the n-width can be calculated theoretically \cite{MAGARILILYAEV200197}. More generally, for functions with its first $(r-1)$ derivatives continuous with respect to parameter variation, and for functions with $C^{\infty}$ dependence, it can be proven that the n-widths are given by $d_n \sim n ^{-r}$ and $d_n \sim e ^{-a n^b}$, respectively \cite{Binev10convergencerates}. These can be thought of as theoretical bounds to the approximation error of a reduced basis. In the case of GWs, they do depend smoothly with respect to parameter variation, thus an optimal basis has {\em asymptotic} exponential convergence with $n$ \cite{PhysRevX.4.031006, Beat2012}; this explains the existence of compact reduced bases of high accuracy (typically machine precision in GW science). 

The task of finding an optimal reduced basis is of combinatorial complexity and unfeasible in practice: all combinations of basis elements must be evaluated in order to find one achieving the optimal n-width. Therefore, computationally cheaper approaches become attractive. An effective one is through a greedy algorithm, which is quasi-optimal in a precise mathematical sense, highly parallelizable, and of linear complexity; 
for more details see \cite{tiglio2021reduced}. 

\subsubsection{Greedy algorithm to build a quasi-optimal reduced basis} 
This approach consists of an iterative procedure, in which the basis is built from a training set $\cal T$, and at each iteration, a new basis function is added to the basis set so that the overall precision is improved. The procedure is said to be greedy in the usual optimization sense: at each iteration, the algorithm chooses the worst element represented from $\cal T$ as a new basis function to be added. The training error of a basis of dimension $n$ (we also refer to it as the {\it greedy error}) is defined as
\begin{equation}
\sigma_{n} := \max_{\lambda} \| h_{\lambda} - \mathcal{P}_n h_{\lambda} \|^{2}.
\end{equation}
In our validation tests, we use an independent validation set of test functions and compute the equivalent of this error; we refer to it as the {\it maximum validation error}. 

From a computational point of view, the cost of enriching a reduced basis with a new element is independent of the dimension of the basis already built, and linear with respect to the size of the training set. 
In terms of accuracy, the algorithm finds a nearly-optimal basis with respect to the Kolmogorov measure: if $d_n$ decays as a power law, so does the greedy error $\sigma_n$, and if $d_n$ decays exponentially with $n$, the same applies to $\sigma_n$~\cite{Devore2012,Buffa2012,Binev10convergencerates}. 

A pseudocode of the greedy procedure is presented in Algorithm~\ref{alg:greedy}, whose main points are explained next: 
\begin{itemize}
\item As input, a training set $\cal T$ of size $m$ is given, along with its associated parameters $\{\lambda_i\}_{i=1}^{m}$, the target maximum training representation error $\epsilon$, and the maximum dimension $n_{max}$ of the basis.

\item First, a function of $\cal T$ is chosen and defined as the first element or seed of the reduced basis (Step 1). Note that for a global basis, this choice is not relevant \cite{Caudill:2011kv}; as we will see, this is very different in hp-greedy. Next, the basis is enriched iteratively with the function of $\cal T$ that is worst represented by an orthogonal projection onto the span of the basis (the corresponding parameter is found in Step 4). 

\item From a practical viewpoint, the different solutions might be almost linearly dependent, resulting in a large conditioning number of the Gram matrix \cite{taylor_1978} used to calculate the projections. Therefore, it is convenient to orthonormalize the solutions to obtain the basis functions. Here, a Gram-Schmidt orthonormalization algorithm \cite{Hoffmann_IMGS} is applied (Steps 5 and 6).

\item The representation error $\sigma_n$ is computed at each iteration (Step 8). 

\item The algorithm ends when $\sigma_n \le  \epsilon$ or $n = n_{max}$, with a reduced basis of dimensionality $n$, built with $h_{\Lambda} = \{h_{\Lambda_i}\}_{i=1}^{n}$, where $\Lambda = \{\Lambda_i \}_{i=1}^{n}$  are referred to as the {\it greedy parameters}.

\item The outputs are the reduced basis $RB$, $\Lambda$, and $\sigma$.
\end{itemize}

\begin{algorithm}[H]
\caption{${\tt GreedyRB}(\lambda, h_{\lambda}, \epsilon, n_{max})$}
\label{alg:greedy}
\begin{algorithmic}[1]
\vskip 10pt
\REQUIRE $\lambda, h_{\lambda}, \epsilon, n_{max}$
\vskip 10pt
\STATE $i=1$, $\sigma = 1$, $\Lambda_1 = \lambda_1$, $RB = \{h_{\Lambda_1}/
\| h_{\Lambda_1} \|  \}$
\WHILE{$\sigma > \epsilon$ and $i < n_{max}$}
\STATE $i=i+1$
\STATE $\Lambda_{i} = \text{argmax}_{ \lambda} \| h_{\lambda} - {\cal P}_{i-1} h_{\lambda} \|^2 $ ~(selection of greedy parameter)  
\STATE $e_{i} = h_{\Lambda_i} - {\cal P}_{i-1} h_{\Lambda_i}$ ~(Gram-Schmidt)

\STATE $e_i = e_i / \| e_i \|$ {\hskip0.725in} (normalization)
\STATE RB = RB $\cup$ $\{e_i\}$
\STATE $\sigma = \max_{ \lambda } \| h_{\lambda} - {\cal P}_{i} h_{\lambda} \|^2$ ~(representation error) 
\ENDWHILE
\vskip 10pt
\ENSURE $RB, \Lambda = ¨ \{\Lambda_{i}\} _{i=1}^{n}, \sigma$
\vskip 10pt
\end{algorithmic} \label{alg:greedy-rb}
\end{algorithm}

\subsection{h-refinement} \label{sec:h}

The terminology \textit{h-refinement} is borrowed from finite differences/elements, where the size of each cell on the mesh is often denoted by $h$. In the context of differential equations, {\it h-refinement} then refers to improving the accuracy of the quantity of interest by decreasing $h$, either by adding more points per domain or by decreasing the size of the latter. We will not elaborate much on this analogy here, it suffices to say that in this context we deal with a domain decomposition in parameter space, which is recursively partitioned and results in a binary tree structure.  

We introduce some notation: 
\begin{eqnarray*}
V &=& \text{parameter space for a given subdomain} \\
D &=& \cup V \, , \text{entire parameter space}  \\
V_1, V_2 &=& \text{partitions of V}\\
\Lambda_{V}  &=& \text{greedy parameters for } V\\
\hat{\Lambda}_{V} &=&  \Lambda_{V}[1]  \, , \text{anchor point for } V \\
\hat{\Lambda}_{V_1}, \hat{\Lambda}_{V_2} &=&   \Lambda_{V }[1]  \,  \text{ or } \Lambda_{V}[2]
\end{eqnarray*}
Each (sub)domain $V \subseteq D$ has an anchor point, which we denote by $\hat{\Lambda}_{V}$. For the domain decomposition or partitioning of V, we assume that the anchor points $\hat{\Lambda}_{V_1}$, $\hat{\Lambda}_{V_2}$ are known and a sampling of V, $\lambda_V$, is given. A {\it proximity function} $d = d (\lambda_1 , \lambda_2 )$, 
$$
d (\lambda_1 , \lambda_2 ) = \| \lambda_1- \lambda_2 \|_2 \, , 
$$
is used to find the anchor point being closest to each parameter of $\lambda_V$. Then, two sets of parameters $\lambda_{V_1}$, $\lambda_{V_2}$ are created, each one with the parameters nearest to one of the two anchor points. In case a point is at the same distance from the two anchor points, it can be arbitrarily assigned to any of the sets, or both. Finally, $\lambda_{V_1}$ and $\lambda_{V_2}$ are returned, representing a sampling of the partitions $V_1$ and $V_2$, respectively. A pseudocode for domain decomposition through a binary partitioning is described in Algorithm~\ref{alg:partition}.

\begin{algorithm}[H]
\caption{ ${\tt Partition}(\lambda_{V}, \hat{\Lambda}_{V_1}, \hat{\Lambda}_{V_2})$ }
\vskip 10pt
\label{alg:partition}
\begin{algorithmic}[1]
\REQUIRE $\lambda_{V}, \hat{\Lambda}_{V_1}, \hat{\Lambda}_{V_2}$
\vskip 10pt 
\STATE $\lambda_{V_1} = \emptyset = \lambda_{V_2}$
\FORALL{$\lambda_i \in \lambda_{V}$}
	\IF{$d(\lambda_i, \hat{\Lambda}_{V_1}) < d(\lambda_i, \hat{\Lambda}_{V_2})$}
		\STATE $\lambda_{V_1} = \lambda_{V_1} \cup \{ \lambda_i \}$
	\ELSIF{$d(\lambda_i, \hat{\Lambda}_{V_1}) > d(\lambda_i, \hat{\Lambda}_{V_2})$}
		\STATE $\lambda_{V_2}= \lambda_{V_2}\cup \{ \lambda_i \}$
\ELSE 
\STATE $\lambda_{V'} = \text{random choice} ([\lambda_{V_1},\lambda_{V_2}])$
\STATE $ \lambda_{V'} = \lambda_{V'}\cup \{ \lambda_i \}$
	\ENDIF
\ENDFOR
\vskip 10pt
\ENSURE $\lambda_{V_1}, \lambda_{V_2}$ 
\end{algorithmic} \label{alg:partition}
\end{algorithm}

As the initial parameter domain $D$ is decomposed, the recursive partitions are structured in a binary tree, in which each node corresponds to a subspace obtained with Algorithm~\ref{alg:partition}. 
The maximum number of levels $\ell $ among all branches of the tree (i.e., its depth) is denoted by ${\ell}_{max}$ , $0\leq \ell \leq {\ell }_{max}$, where $\ell= 0$  represents the case with no partitioning at all (i.e., the standard reduced basis approach). 
Each node at level $\ell$ is labeled by a series of indices $B_{\ell}$ as  
$$
B_{\ell} = (0,i_2, \ldots, i_{\ell})\, , \text{with} \,   i_j= \{0,1\} \, , 
$$
where, by convention, $i=0$ for the left leave and $i=1$ for the right one. For example, the root ($\ell = 0$) comprises the whole parameter domain and is labeled by 
$$
B_{0} = (0,) \, , 
$$
and its two children ($\ell = 1$) by 
$$
B_{1} = (0,0) \, \text{or} \, (0,1)\, .
$$

\begin{figure}[ht]
    \centering
    \includegraphics[width=0.4\linewidth]{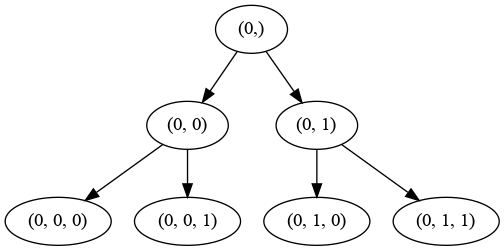}
    \caption{Tree representing a partitioning of the parameter space. The index notation for the nodes of a tree of depth ${\ell}_{max}=2$ is shown.}
    \label{fig:tree}
\end{figure}

Figure~\ref{fig:tree} gives an example of the notation for a tree with ${\ell}_{max}=2$, where all leaves reach the maximum allowed depth . 
\subsection{hp-greedy refinement} \label{sec:alg}

This approach combines {\it h-refinement} and {\it p-refinement} within the greedy reduced basis framework. In order to decide whether to partition a given parameter domain, a reduced basis is built, with an accuracy threshold $\epsilon$ and stopping criteria of $n_{max}$ basis elements {\em per subdomain}, as in Algorithm~\ref{alg:greedy-rb}. If the accuracy threshold is not reached, then the anchor point of the domain to be split is used as the anchor point of its left children, and the second greedy parameter of the reduced basis of the (parent) domain as the anchor point for the right children. Then, a binary domain decomposition is performed, as described in Algorithm~\ref{alg:partition}. Next, a reduced basis for each child is built. 
Note that the reduced basis from a parent node is {\em not} reused when building the bases for each of its children.
The children domains where $\epsilon$ is not reached within $n_{max}$ greedy iterations are further split into two more children according to Algorithm~\ref{alg:partition}. This adaptive process continues until the accuracy threshold $\epsilon$ is achieved, the depth of the partition $l$ reaches $l_{max}$, or the number of training set parameters is exhausted. If the latter happens, it means that the accuracy threshold cannot be reached with the prescribed $n_{max}$ stopping criteria. The maximum allowed depth of the tree, $l_{max}$ , is referred to as {\it early stopping} in machine learning.

A pseudocode for this hp-greedy refinement \cite{Eftang:2010} approach is presented in Algorithm~\ref{alg:hp-greedy}. 

\begin{algorithm}[H]
\caption{${\tt hpGreedy}(\lambda_{V}, h_{\lambda_{V}}, \epsilon, n_{max}, \ell, {\ell }_{max}, B_{\ell})$}
\label{alg:hp-greedy}
\begin{algorithmic}[1]

\vskip 10pt
\REQUIRE  $\lambda_{V}, h_{\lambda_{V}}, \epsilon, n_{max}, \ell, {\ell}_{max}, B_{\ell}$

\vskip 10pt

\STATE $RB, \Lambda_{V},\sigma = {\tt GreedyRB}( \lambda_{V}, h_{\lambda_{V}}, \epsilon, n_{max})$

\vskip 10pt
\IF{$\sigma > \epsilon \algword{and} \ell < {\ell}_{max}$}
	\STATE $\hat{\Lambda}_1 = \Lambda_{V}[1]$
	\STATE $\hat{\Lambda}_2 = \Lambda_{V}[2]$
	\STATE $\lambda_{V_1}, \lambda_{V_2} = {\tt Partition}(\lambda_{V}, \hat{\Lambda}_1, \hat{\Lambda}_2) $
	\STATE $out_1 = {\tt hpGreedy}(\lambda_{V_1}, h_{\lambda_{V_1}}, \epsilon, n_{max}, \ell + 1, {\ell}_{max}, (B_{\ell}, 0))$
	\STATE $out_2 = {\tt hpGreedy}(\lambda_{V_2} , h_{\lambda_{V_2}}, \epsilon, n_{max}, \ell + 1, {\ell}_{max}, (B_{\ell}, 1))$
	\STATE $out = out_1 \cup  out_2$
\ELSE
	\STATE $out = \{ (RB, \Lambda_V, h_{\Lambda_V}, B_{\ell}) \}$
\ENDIF

\vskip 10pt
\ENSURE out
\vskip 10pt
\end{algorithmic}
\end{algorithm}
The first greedy parameter of the global basis (the seed of the algorithm), built for the first partition of the entire domain $D$, can {\em in principle} be chosen arbitrarily (we will see that it does have an impact on the accuracy of the resulting bases). Taking into account that the partitioning is carried out with the first two greedy parameters of the reduced basis (Step 5 in Algorithm~\ref{alg:hp-greedy}), we notice that the seed of the algorithm determines the first partition, and thus, it also conditions the successive partitions and the reduced bases associated with those partitions. As we discuss in Section \ref{sec:gws}, the seed is a relevant hyperparameter of the algorithm.

There is no rigorous rule for choosing $n_{max}$, $l_{max}$ and the seed. From a machine learning perspective, a possible approach to find them is through hyperparameter optimization, as they can be seen as hyperparameters of the algorithm in the sense of being parameters whose values are set before the learning process begins.

The rationale of hp-greedy is that if the greedy error is decaying slowly,
the number of greedy iterations in the domain to be split gets too large, then the domain is partitioned (i.e., refined).  The notion of a slowly decaying error is problem-dependent. 

\section{A toy model application} \label{sec:toy}
\begin{figure}[ht]
    \centering
    \includegraphics[width=0.5\linewidth]{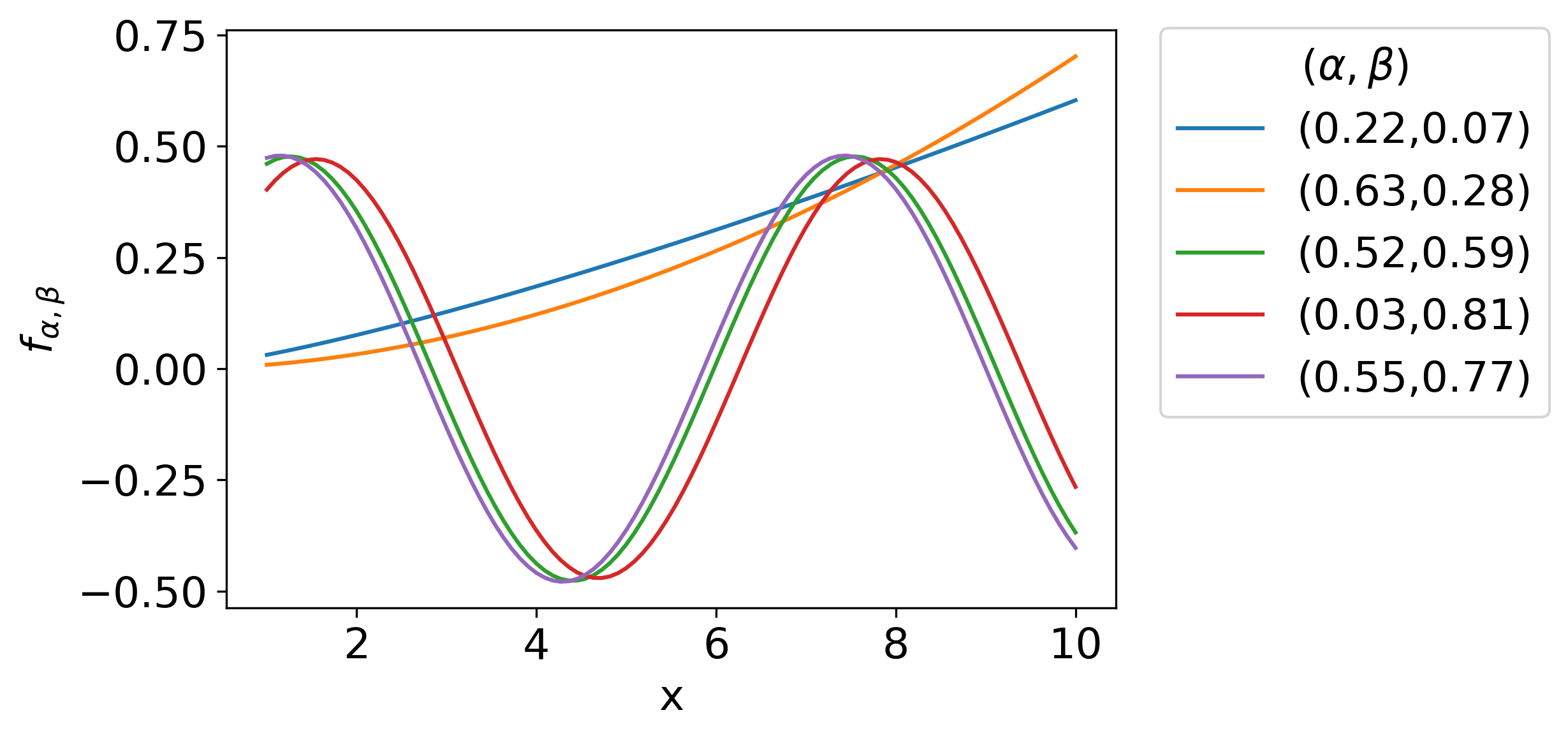}
    \caption{A random sample of normalized functions $f_{\alpha,\beta}$ for our toy model.  As it can be seen, the functions have a very different structure and there is a clear discontinuity in the parameter space.}
    \label{fig:toy_model_f}
\end{figure}

We illustrate how hp-greedy reduced basis works for a toy model of functions that we artificially constructed with an intended strong discontinuity in the parameter space $D$. The hp-greedy procedure was run several times for a grid of hyperparameters $({\ell}_{max},n_{max})$, a fixed seed and a greedy tolerance $\epsilon=10^{-10}$.  
For visualization purposes, we chose $D$ to be two-dimensional:
$$
D:=[0,1]\times[0,1] \, , 
$$
labeled by a tuple $(\alpha,\beta)$. Our chosen parametrized functions are of the form $f_{\alpha,\beta}(x): \mathbb{R} \rightarrow \mathbb{R} $ with  $x \in [0,10]$ and

\begin{equation}
f_{\alpha,\beta}(x):=
     \begin{cases}
       \text{$x^{1+\beta+\alpha}$,}   & \text{$ \beta < 1/2$}\\
       \text{$\sin(\beta  \alpha + x)$,} & \text{$1/2 \leq \beta $} \\
     \end{cases}
\end{equation}
After this choice, we normalized the functions so that
$$
\left \| f_{\alpha,\beta} (\cdot )\right \|^2 := \int_0^{10} f^2_{\alpha,\beta}(x) dx = 1 \quad \forall \, (\alpha, \beta) \,  , 
$$
to place emphasis on their structure rather than on their size. Examples of $f_{\alpha,\beta}$ for different parameter values are shown in Figure~\ref{fig:toy_model_f}.

We sampled a training set with $100$ equispaced points per parameter dimension. That is, $100 \times 100$ numerical values of $(\alpha,\beta)$ were chosen and used by hp-greedy to subdivide each domain. For the validation set, we used $101$ different (from those of the training set) equispaced points per parameter dimension.

To build hp-greedy models we used the open-source Python package Arby~\cite{arby} to obtain a reduced basis for a given training set of waveforms (Algorithm~\ref{alg:greedy}). In addition, special-purpose code was written to deal with the partitioning of the domain (Algorithm~\ref{alg:partition}) using hp-greedy (Algorithm~\ref{alg:hp-greedy}).

\subsection{Domain partitioning} \label{sec:domains}

By design, the algorithm is expected to automatically identify regions in the parameter space where functions have dissimilar structures, and then recursively divide them into subdomains. 
Each subdomain has its own reduced basis, with at most $n_{max}$ elements, and is partitioned until a given threshold $\epsilon$ is reached or the depth of the tree ${\ell}$ is equal to ${\ell}_{max}$.

\begin{figure}[ht]
    \centering
    \includegraphics[width=0.45\linewidth]{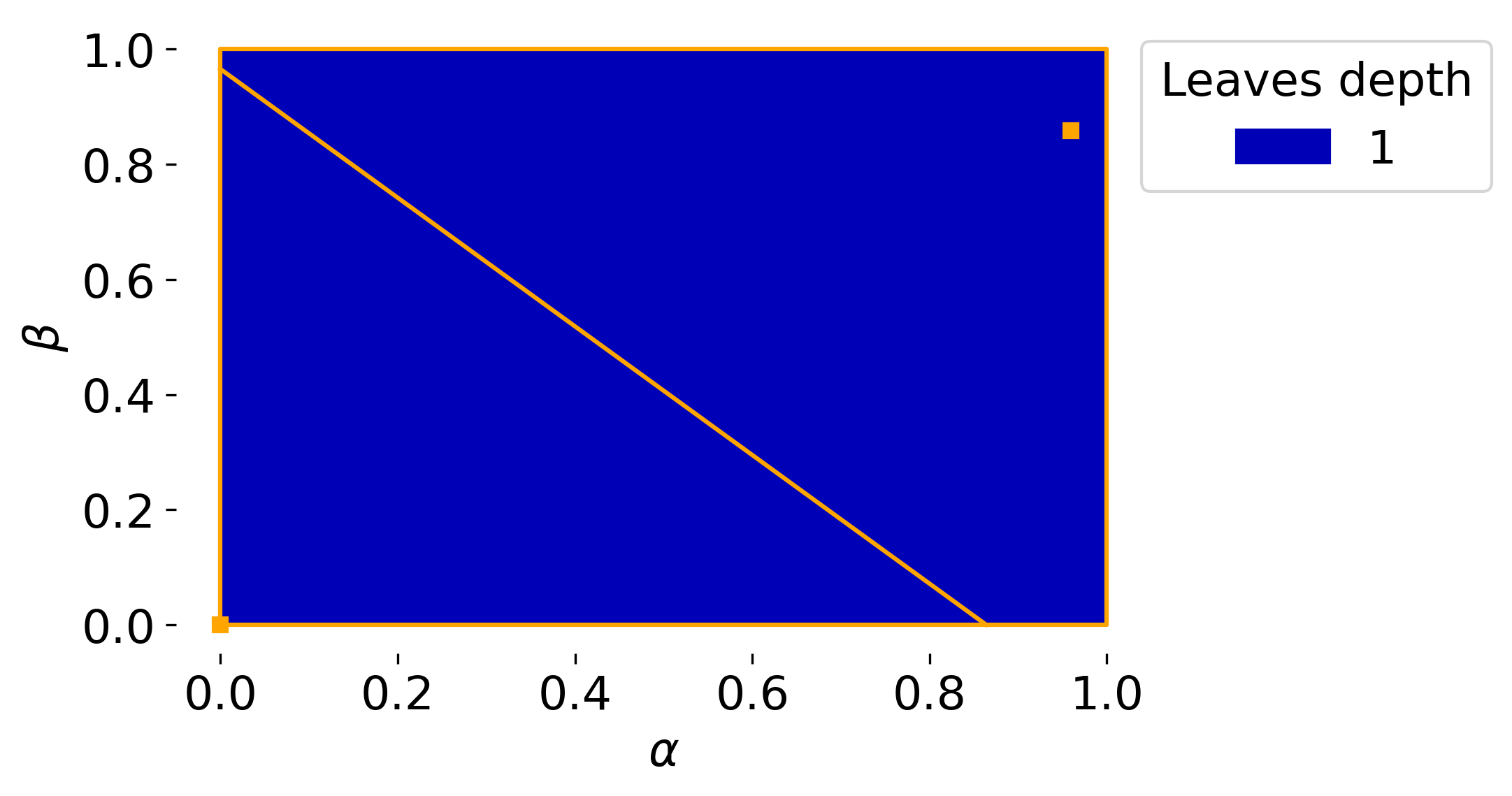}
    \includegraphics[width=0.45\linewidth]{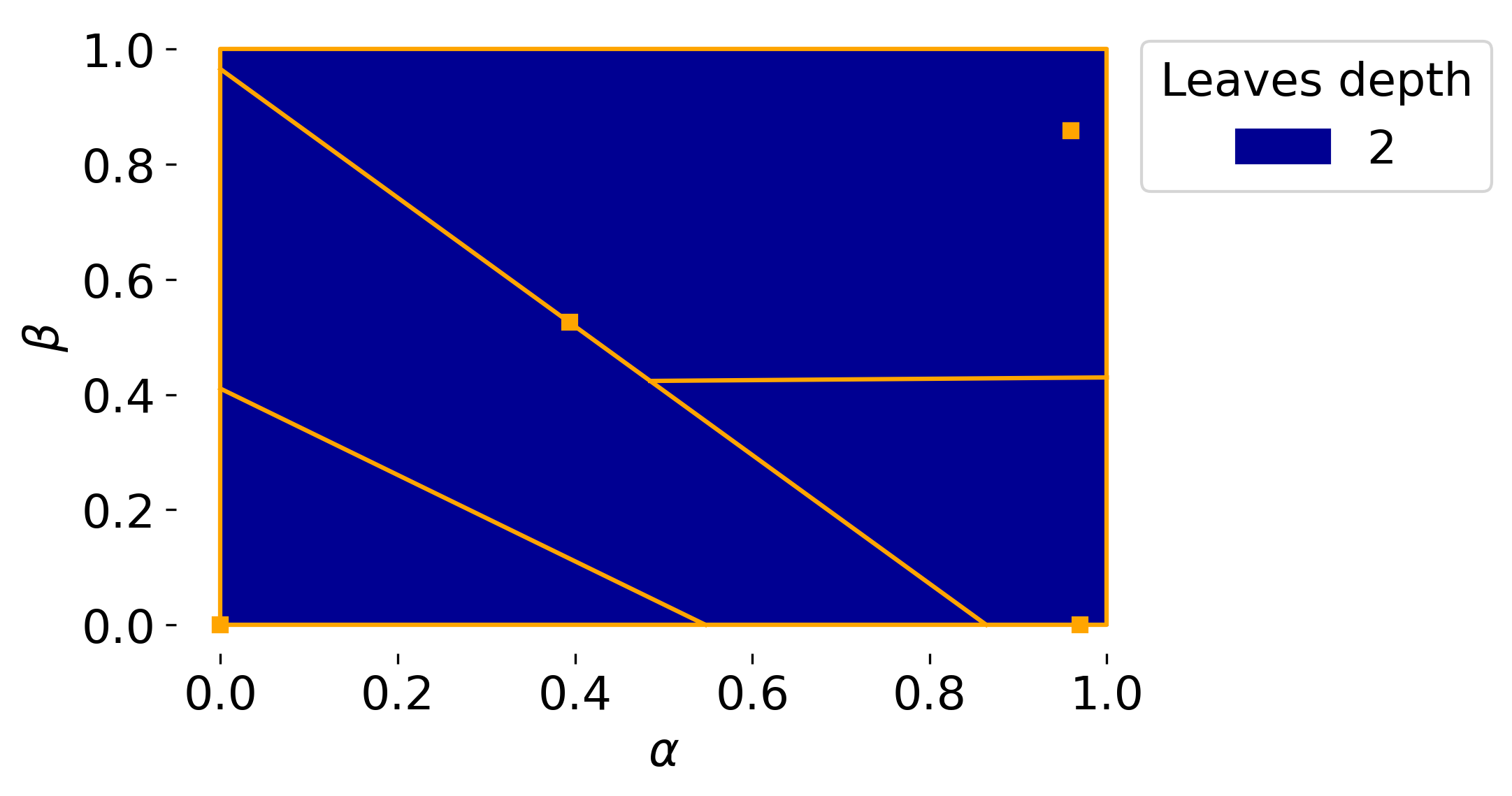}
    \includegraphics[width=0.45\linewidth]{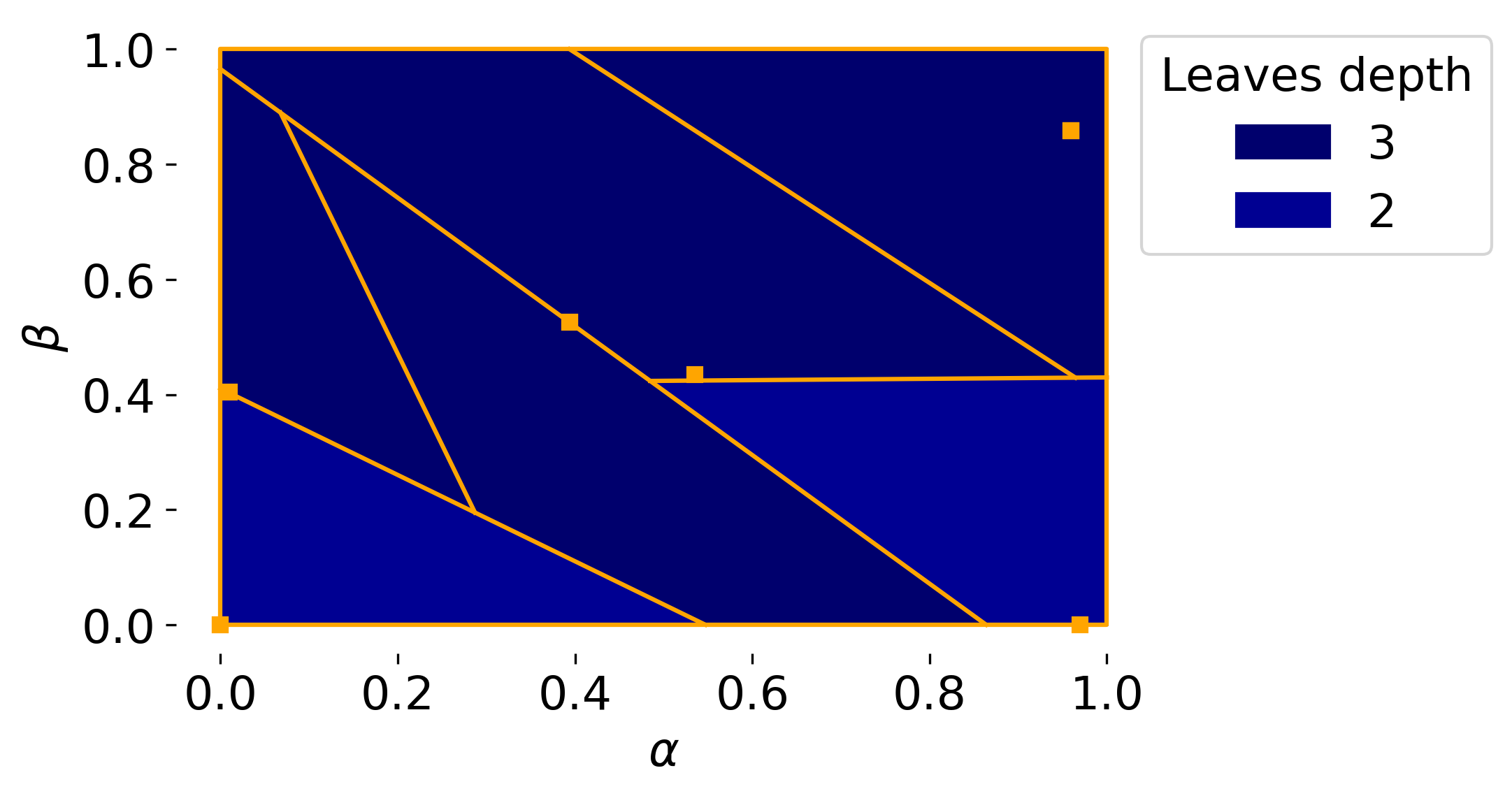}
    \includegraphics[width=0.45\linewidth]{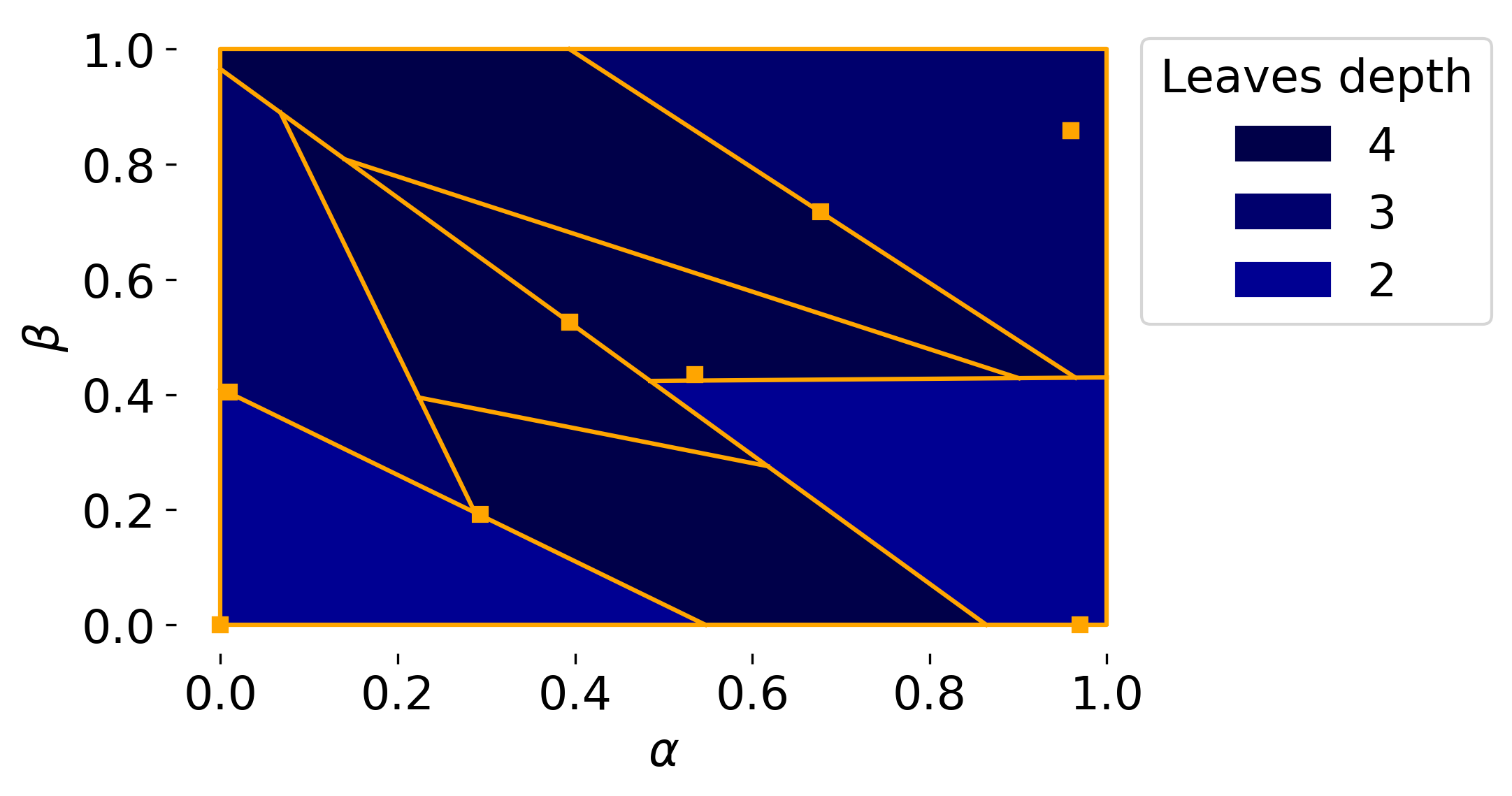}
    \includegraphics[width=0.45\linewidth]{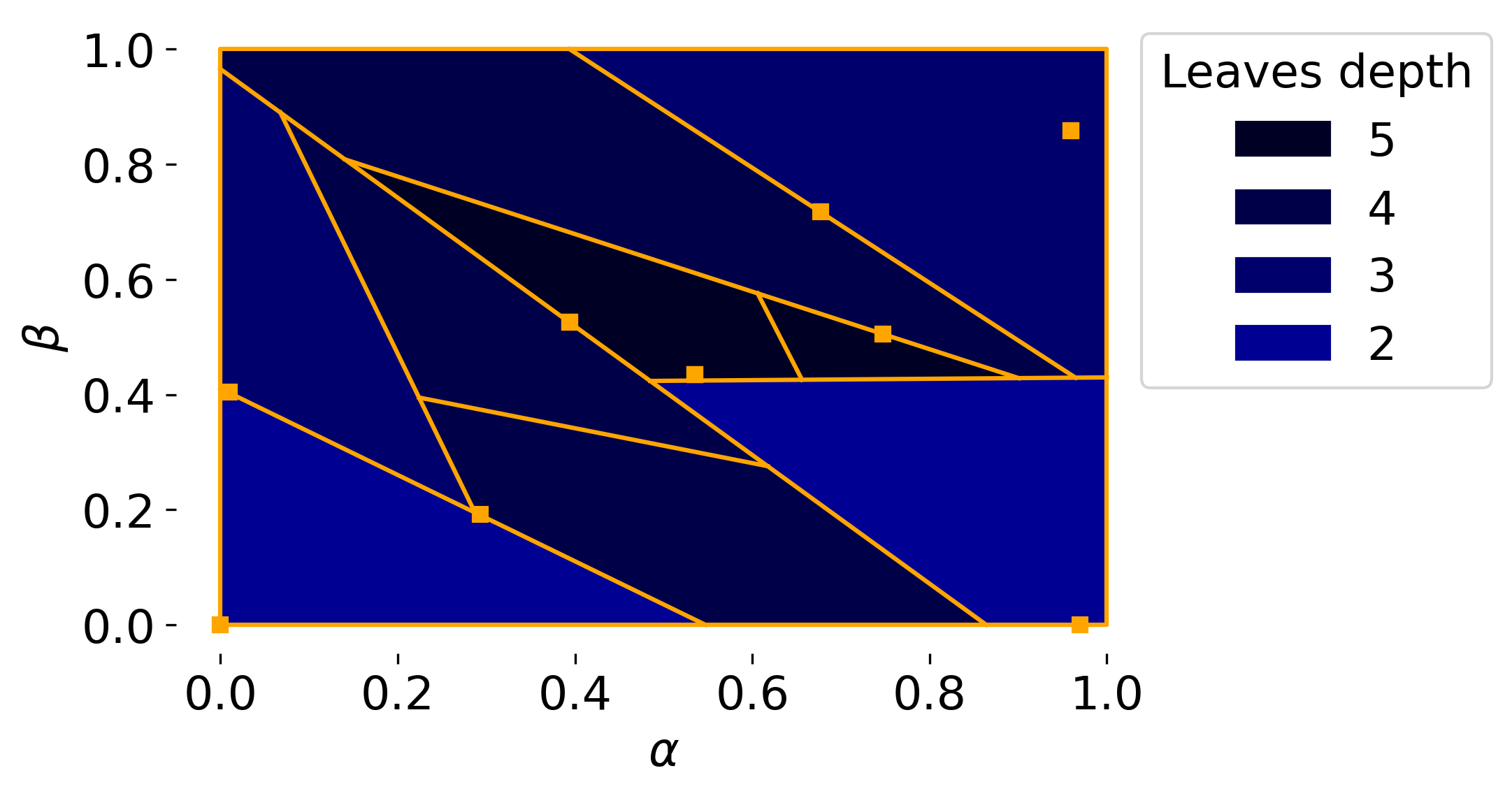}
    \caption{Successive partitions performed by hp-greedy for the toy model, with ${\ell}_{max} = n_{max} = 5$. The {\it leaves depth} is the value of $\ell $ of that subdomain. The yellow squares denote the anchor points obtained by the hp-greedy algorithm to partition the parameter space.}
\label{fig:toy_model}
\end{figure}

To illustrate hp-greedy in a specific example, we analyzed the successive partitions obtained at each step of the algorithm, with ${\ell}_{max} = n_{max} = 5$ and seed $(\alpha,\beta)=(0,0)$, as exemplified in Figure~\ref{fig:toy_model}.    
A darker color of the partition indicates that the algorithm needed more steps to describe that subdomain; that is, the corresponding leave in the tree is deeper (has a larger value of $\ell $) than those of the partitions with a lighter color. 

The upper left plot corresponds to the first iteration of hp-greedy, in which the domain is divided into two parts, using the first two greedy parameters of the global RB as anchor points. The next plot (on the right) shows the second domain decomposition, in which each subspace (from the first iteration) is partitioned using the first two greedy parameters of its associated RB.
The plots below show the partitionings of the following iterations. In this case the algorithm had to do more partitions in the region around the discontinuity, because the change of the functions with parameter variation is larger than the places where there is no discontinuity and the variation of the functions is smooth.

\begin{figure}[ht]
    \centering
    \includegraphics[width=0.6\linewidth]{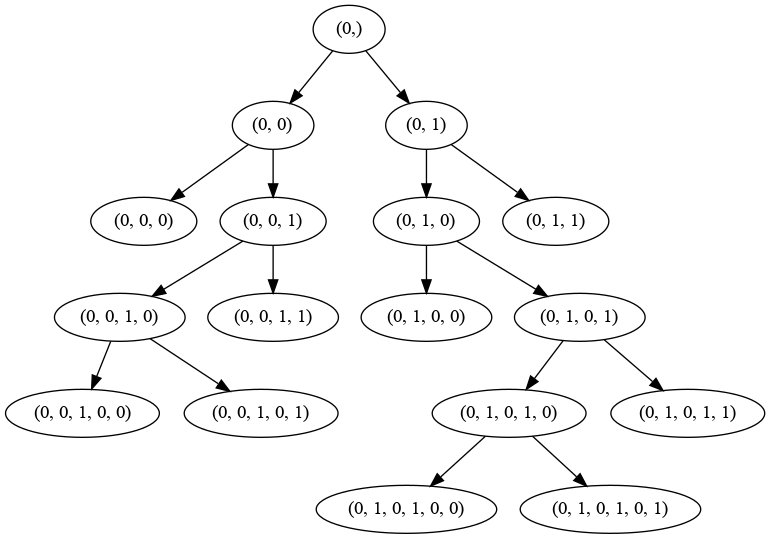}
    \caption{Tree produced by the hp-greedy algorithm for the toy model with ${\ell }_{max} = n_{max}=5$. For leaves with depth $\ell < {\ell }_{max}$, hp-greedy stopped when reaching the threshold training error ${\epsilon} = 10^{-10}$. }
    \label{fig:toy_model_tree}
\end{figure}

Figure~\ref{fig:toy_model_tree} shows the resulting tree structure for this toy model. The two deepest leaves, with $\ell =5$, $(0,1,0,1,0,0)$ and $(0,1,0,1,0,1)$, contain the region of the discontinuity, which is harder to represent when compared to a domain with no discontinuity. These two nodes are associated with the subspaces in the center of the parameter space, which are black-colored in the fifth plot of Figure~\ref{fig:toy_model}.

\begin{figure}[ht]
    \centering
    \includegraphics[width=0.45\linewidth]{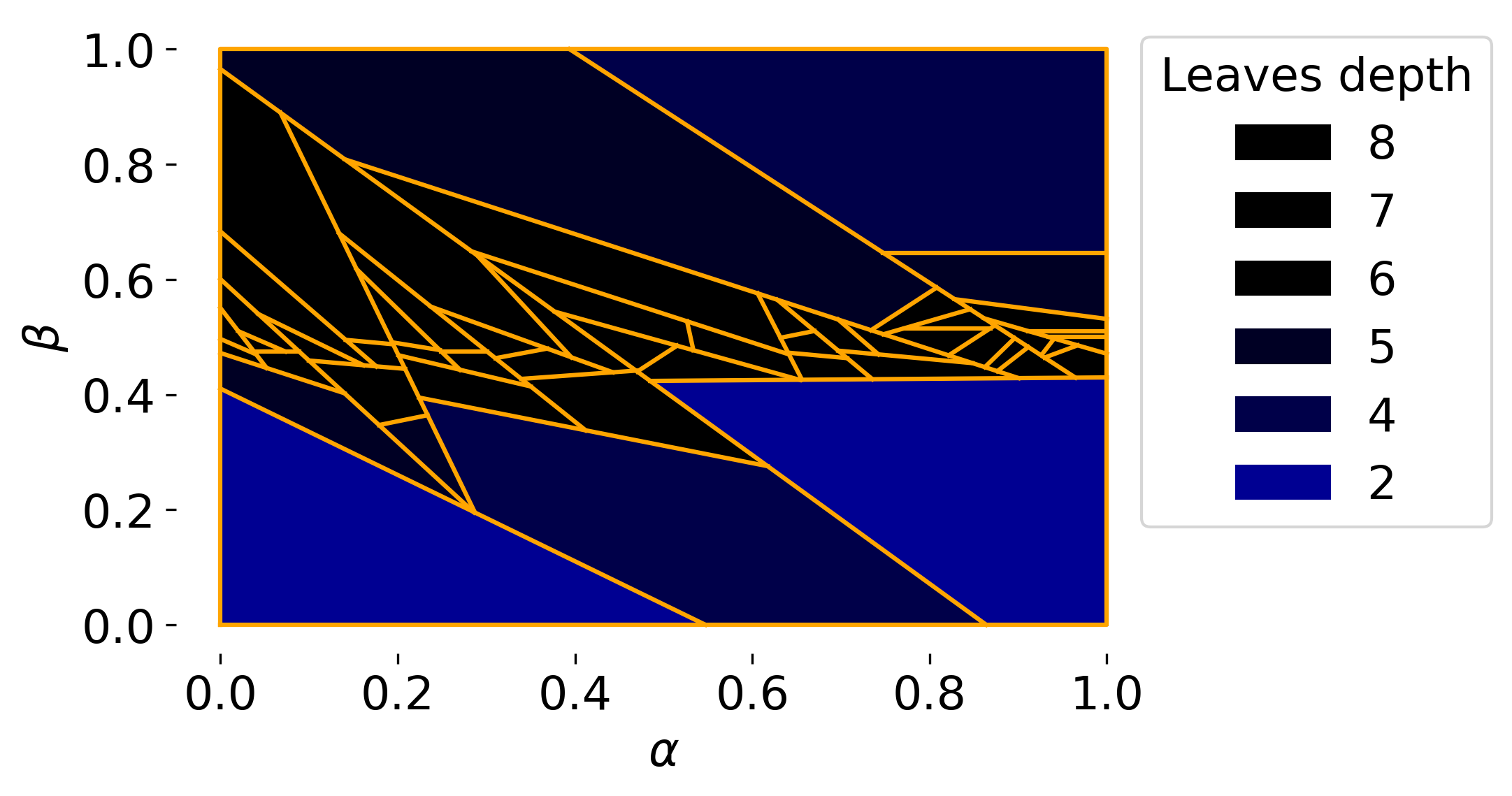}
    \caption{Partitions performed by hp-greedy for the toy model, with ${\ell}_{max} = 8$ and $n_{max} = 4$. Notice that hp-greedy needed to refine notably more close to where the discontinuity is located, at $\beta = 0.5$.}
\label{fig:toy_model_disc}
\end{figure}

To assess the behavior of the partitioning in a more extreme case, we let the algorithm keep partitioning with a larger $l_{max}$: we set ${\ell}_{max} = 8 $ and $n_{max} = 4$. The result is shown in Figure~\ref{fig:toy_model_disc}.  In this case, it can also be seen that the algorithm performed more partitions in the region of the discontinuity, almost ``detecting" the discontinuity. Furthermore, the partitioning stopped earlier in those subspaces with no discontinuity.

\subsection{Convergence} \label{sec:conv}
Figure~\ref{fig:toy_model2} presents some results for different combinations of hyperparameters $(n_{max}, {\ell}_{max})$ using validation data. 
The first aspect to emphasize here is the observed spectral convergence: for a fixed ${\ell}_{max}$, the validation errors decay exponentially as a function of $n_{max}$, even with a discontinuity in the parameter space. It is unclear to us why this is so even for ${\ell}_{max}=0$, (i.e. no partitioning) and without evidence of Gibb's phenomenon \cite{Jan-RB,hesthaven2007spectral}; we speculate that this effect should appear at very high resolutions, below our chosen threshold $\epsilon=10^{-10}$. 

We now focus on the differences between a global basis, {${\ell}_{max}=0$,  and partitioning, ${\ell}_{max}>0$, by comparing them for a given maximum dimensionality $n_{max}$ for each basis \footnote{We point out that the {\em total} number of basis elements is, in general, larger if there is a partition of the parameter space (${\ell}_{max}>0$) because there are more basis with the same constraint: a dimension less or equal than $n_{max}$.}.  
This comparison can be qualitatively seen by fixing $n_{max}$ in the left panel of Figure~\ref{fig:toy_model2}. Except for cases of very low dimensionality and poor accuracy ($ \epsilon < 10^{-2}$, $n_{max}<3$), we notice that increasing ${\ell}_{max}$ significantly improves the maximum validation error. 
As an example, for $n_{max}=5$, from a global basis ${\ell}_{max}=0$ to, say, ${\ell}_{max}=5$, there are around four orders of magnitude improvements in the error.

We finally focus on the value of $n_{max}$ needed to achieve a given representation error for different values of $l_{max}$. This analysis is appealing because a model with a lower $n_{max}$ and the same or better accuracy can yield faster representations for the same precision. This happens because the dimensionality of the basis is a key point  when evaluating a representation or surrogate, and it could be a way to accelerate statistical inference, as discussed in Section~\ref{sec:com}. In our study, it can be seen that a model with partitioning can have a lower $n_{max}$ than a model with no partitioning and the same or lower error. For example, for an error of $10^{-6}$, models with $l_{max}= 2, 3, 4 ,5$ have a comparable or lower error than the case without partitioning (see Figure~\ref{fig:toy_model2}).

\subsection{Overfitting} \label{sec:overf}
Overfitting is a well-known behavior in machine learning, which entails that the learning process does not improve indefinitely when using more complex models; in fact, the errors might become worse~\cite{James2013}. This effect becomes evident when the training error decreases, but the opposite happens with the validation errors. 

As the successive partitions of hp-greedy can be structured in a tree, overfitting in our approach can be related to the standard overfitting pattern of decision trees. If the maximum depth ${\ell}_{max}$ is large enough, the training data can be very well represented, but using validation data will likely show overfitting. In other words, there is a tradeoff between accuracy and tree depth when training while avoiding overfitting. Hyperparameter optimization can be approached in a number of ways to deal with this tradeoff, this aspect is left to future work.

In our numerical experiments for the toy model, we found that --as expected-- for certain, but not all, values of $n_{max}$, larger values of ${\ell}_{max}$ lead to models with higher accuracy models for the training data; however, after a certain value, the maximum {\em validation} errors start increasing, up to orders of magnitude. In the left panel of Figure~\ref{fig:overfitting_nmax=3} we show how overfitting takes place for $n_{max} = 3$: it starts at ${\ell}_{max} = 8$ and the difference between maximum training and validation error gets larger than three orders of magnitude. Nonetheless, this pattern of overfitting did not appear in our experiments for all values of $n_{max}$. The right panel of Figure~\ref{fig:overfitting_nmax=3} shows the behavior of the validation error for multiple values of $(n_{max}, \ell_{max})$. It can be seen that for $n_{max} > 4$ an increase in ${\ell}_{max}$ does not necessarily result in overfitting.

 \begin{figure}[ht]
    \centering 
    \includegraphics[width=0.4\linewidth]{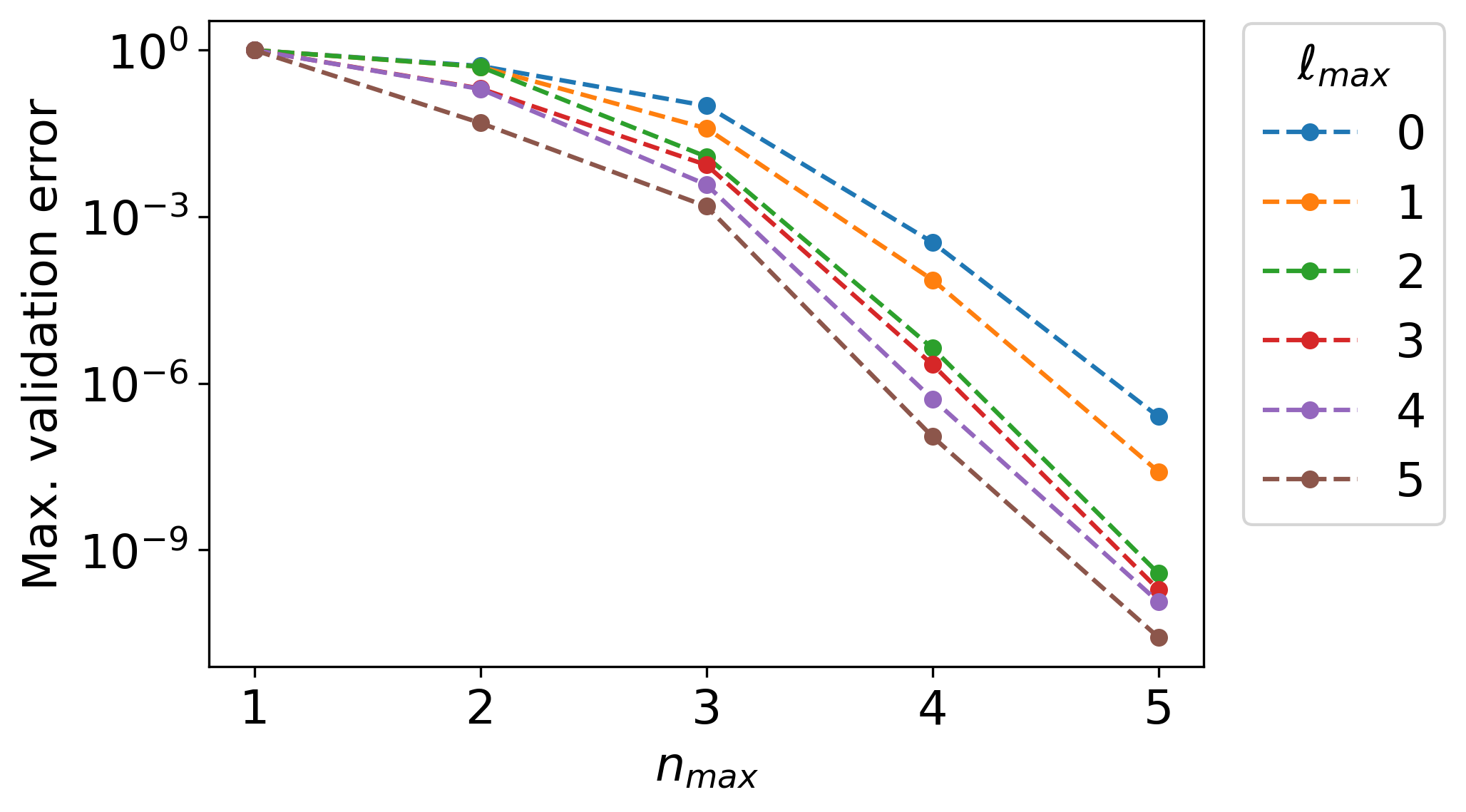}

    \caption{Maximum validation errors for different values of $n_{max}$  and ${\ell}_{max}$ for the toy model. The dashed blue curve shows the results using a global basis (no refinement), while the other curves are based on hp-refinement. }
    \label{fig:toy_model2}
\end{figure}

 \begin{figure}[ht]
    \centering 
\hspace{5mm}%
   \includegraphics[width=0.45\linewidth]{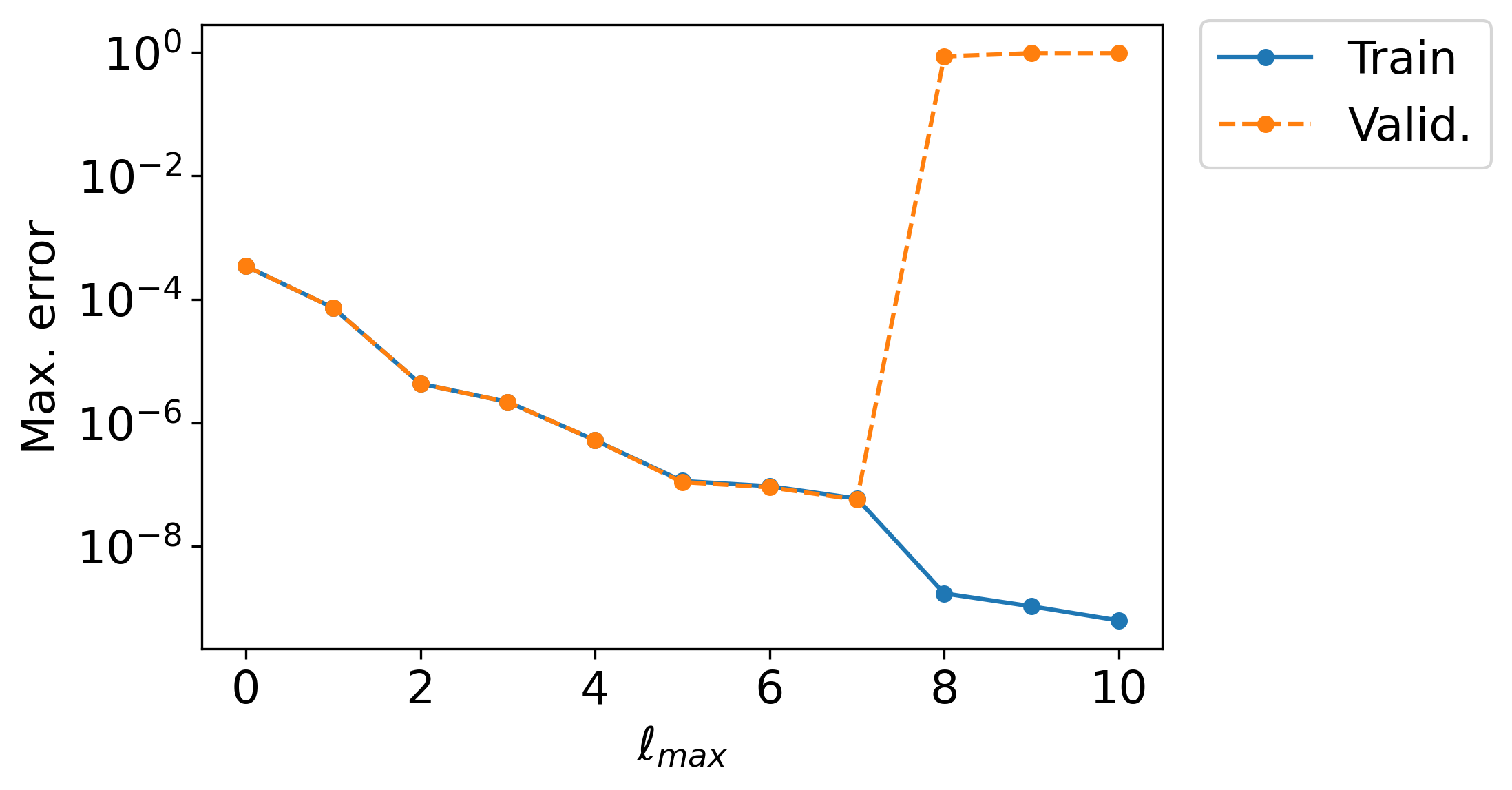}
\includegraphics[width=0.43\linewidth]{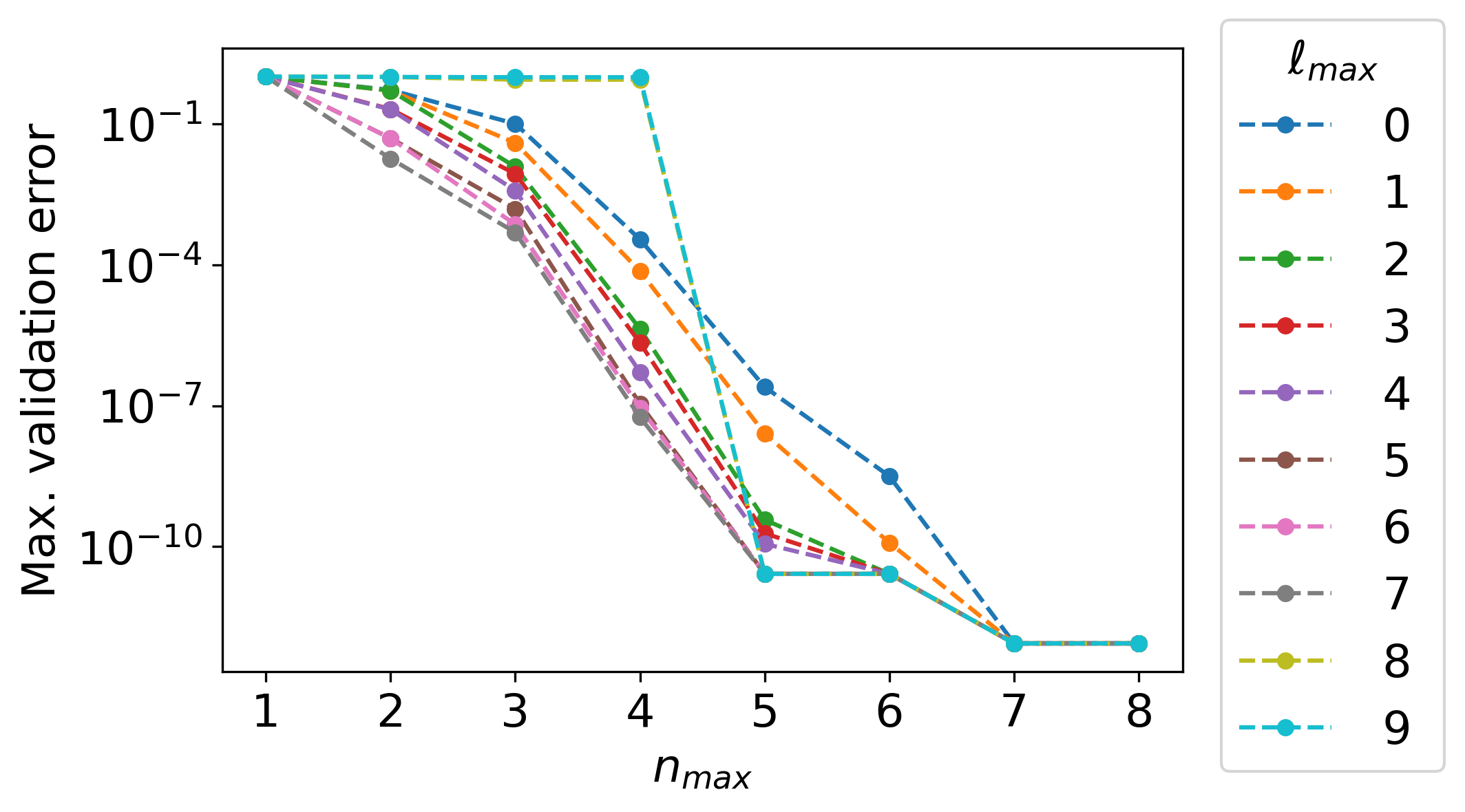}
    \caption{\textbf{Left:} Training and validation errors for the toy model of Section~ \ref{sec:toy} with $n_{max}=3$. Overfitting is present, starting at ${\ell}_{max} =8$.
\textbf{Right:} Maximum validation errors for different values of $n_{max}$  and ${\ell}_{max}$ for the toy model. Overfitting is present for $n_{max} \leq 4$ when increasing ${\ell}_{max}$.
} 
    \label{fig:overfitting_nmax=3}
\end{figure}

\section{Gravitational waves from spinning black hole collisions} \label{sec:gws}
We discuss the results of hp-greedy refinement using the physical setup of the GWs emitted by the collision of two spinning, non-precessing binary black holes initially in quasi-circular orbit. We use the hybrid numerical relativity (NR) and post-Newtonian (PN) surrogate NRHybSur3dq8 \cite{PhysRevD.99.064045} as starting point.

Each waveform $h$  is represented by a complex time series, $h = h_+ - i h_{\times}$, where $h_+$ and $h_{\times}$ are the two polarizations of the gravitational wave. The time domain used in Ref.~\cite{PhysRevD.99.064045} is $[-5.4\times10^{8},135]M$, the long time interval explained by the use of PN  approximations at early times, where $t=0$ represents the peak of amplitude of the waveforms and $M$ is the total mass of the binary system. A flat noise curve, $l \leq 4$ and $(5,5)$ angular modes except for $(4,1)$ or $(4,0)$ were used in the construction of the surrogate. 

In the late inspiral part of the waveforms, starting at $-3500M$ before the peak, the surrogate reproduces waveforms with mismatches $\lesssim 3 \times 10^{-4}$; where the latter are evaluated  computing out-of-sample errors, randomly dividing  the 104 training waveforms into groups of $\sim$5 waveforms each and doing a cross-validation study.  The errors are well within the truncation error of the NR simulations, which are computed calculating the mismatch between the two highest available resolutions of each NR waveform.
\subsection{Datasets} \label{sec:datasets}

In Ref.~\cite{PhysRevD.99.064045} the surrogate NRHybSur3dq8 was built in the parameter range of mass ratios $1 \leq q \leq 8$ and dimensionless spins $-0.8 \leq \chi_{1z} , \chi_{2z}  \leq 0.8$. In this paper, we use those same ranges but only the dominant angular mode $\ell =m =2$ for simplicity and the sake of illustrating hp-greedy. Furthermore, in order to speed up our numerical experiments, we sampled NRHybSur3dq8 waveforms in the late inspiral part and merger regimes $t\in [-3000,130]M$, with $\Delta t = 0.1M$. We also normalized the waveforms with respect to the $L_{2}$ norm (Equation \ref{eqn:l2norm}), to put emphasis on structure/shape, instead of size/amplitude.

We studied three different cases, namely: 

\begin{itemize}
\item 1D: No spin, the only free parameter is $q=m_1/m_2$, due to the scale invariance of GR.
\item 2D: Two aligned spins with the same magnitude are added to the 1D case, meaning that $\chi_{1z}  = \chi_{2z}$.
\item 3D: The two spins are allowed to vary, but independently: in general, $\chi_{1z}  \neq  \chi_{2z}$.
\end{itemize}

Regarding training and validation sets, in the 1D case we generated two different sets of 500 waves to train and validate. For the 2D and 3D cases, we used 3,000 waves to train and 1,000 waves to validate.

 \begin{figure}[ht]
    \centering 
   \includegraphics[width=0.93\linewidth]{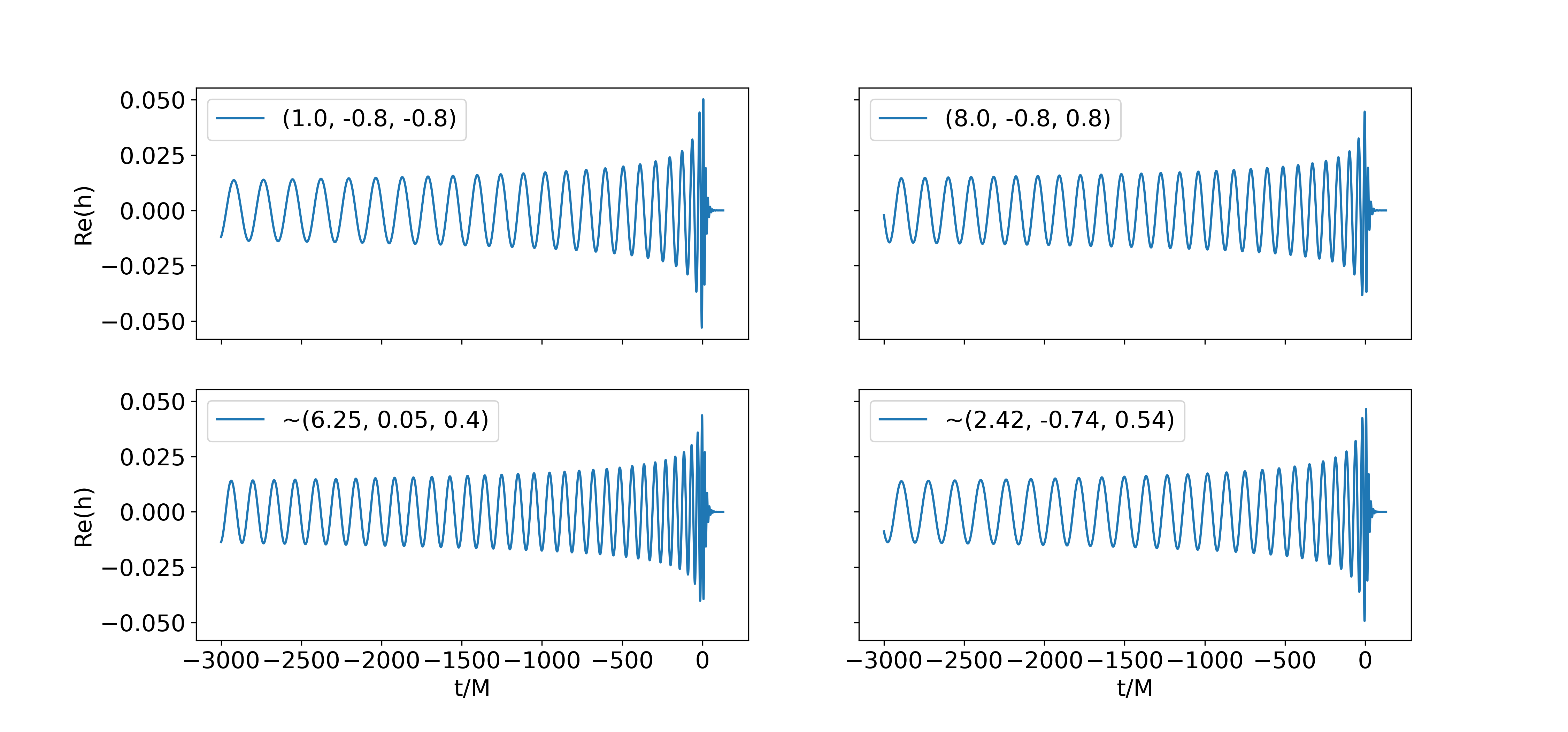}
    \caption{Visualization of the real part of training waveforms. Each one has a different associated parameter value for the tuple $(q, \chi_{1z}, \chi_{2z})$. }
    \label{fig:visual}
\end{figure}

\subsection{hp-greedy refinement, dependence on seed} \label{sec:hyperparameters} 

In our numerical experiments we set the greedy tolerance at double precision, $\epsilon = 10^ {-16}$. As explained in Section~\ref{sec:alg}, a seed is used as the anchor point for hp-greedy to initialize the whole algorithm and build local reduced bases. For a fixed seed, we built a number of hp-greedy bases with different values of $(n_{max}, l_{max})$ and manually chose, for each $n_{max}$, the $l_{max}$ value leading to the highest accuracy multidomain representation. Since overfitting appears for large values of $l_{max}$, as discussed in Section \ref{sec:overf}, our 1D explorations were restricted to $0 \leq {\ell}_{max} \leq 9$, and $0 \leq  {\ell}_{max} \leq 5$ in 2D and 3D. 

With respect to the algorithm seed, it has been consistently found~\cite{Caudill:2011kv,Field:2012if} through numerical experiments that for global reduced bases, its choice is irrelevant because the greedy algorithm performs a global optimization --see for example Figure 1 of~\cite{Caudill:2011kv}. Interestingly, for hp-greedy and local bases, we found that the seed choice is highly relevant, and the accuracy of the resulting bases does depend on its choice. This situation is exemplified in Figure~\ref{fig:seeds}; for example, in the 2D case there are differences of up to three orders of magnitude in the error when varying the seed.  In fact, it is possible that extensive seed searches might reveal larger differences. Therefore, for hp-greedy the seed should be taken as another hyperparameter of relevance, which is one of the main findings of this work.

In Figures \ref{fig:gw_1d_part} and \ref{fig:gw_2d_part} we show the domain decomposition obtained by hp-greedy for the parameter space in 1D and 2D, respectively, for different seed choices. It can be qualitatively seen that there is a significant impact on the partitioning of the domain.

We observed that almost all the trees of the resulting models are balanced and with depth $\ell = {\ell}_{max}$, unlike the toy model of Section~\ref{sec:toy}. This means that in our numerical experiments hp-greedy did not find specific regions where more refinement was needed. Examples of this behavior are shown in Figures \ref{fig:gw_1d_part} and \ref{fig:gw_2d_part}. 

 \begin{figure}[ht]
    \centering 
\hspace{-2mm}%
\includegraphics[width=0.49\linewidth]{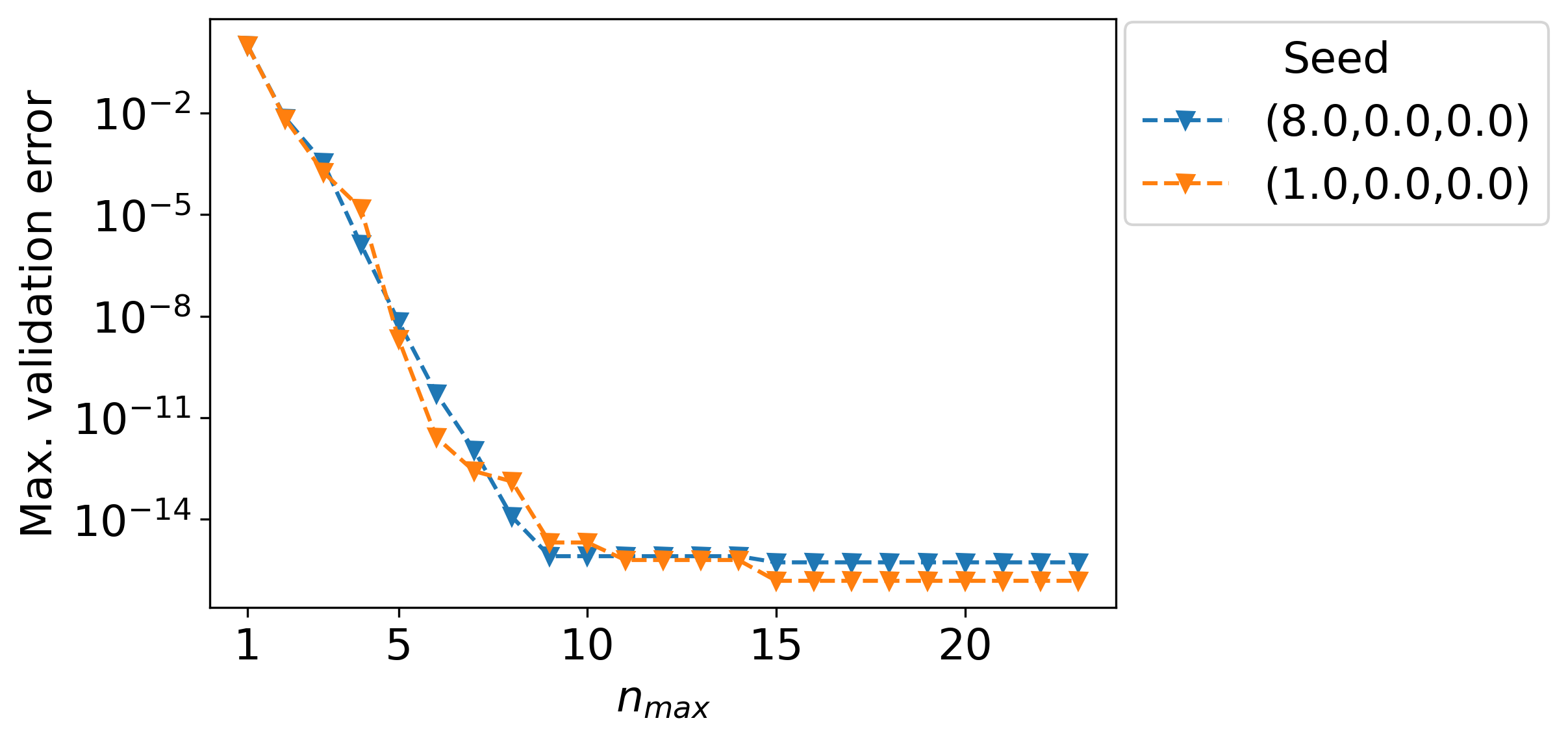}
\hspace{2mm}%
\includegraphics[width=0.49\linewidth]{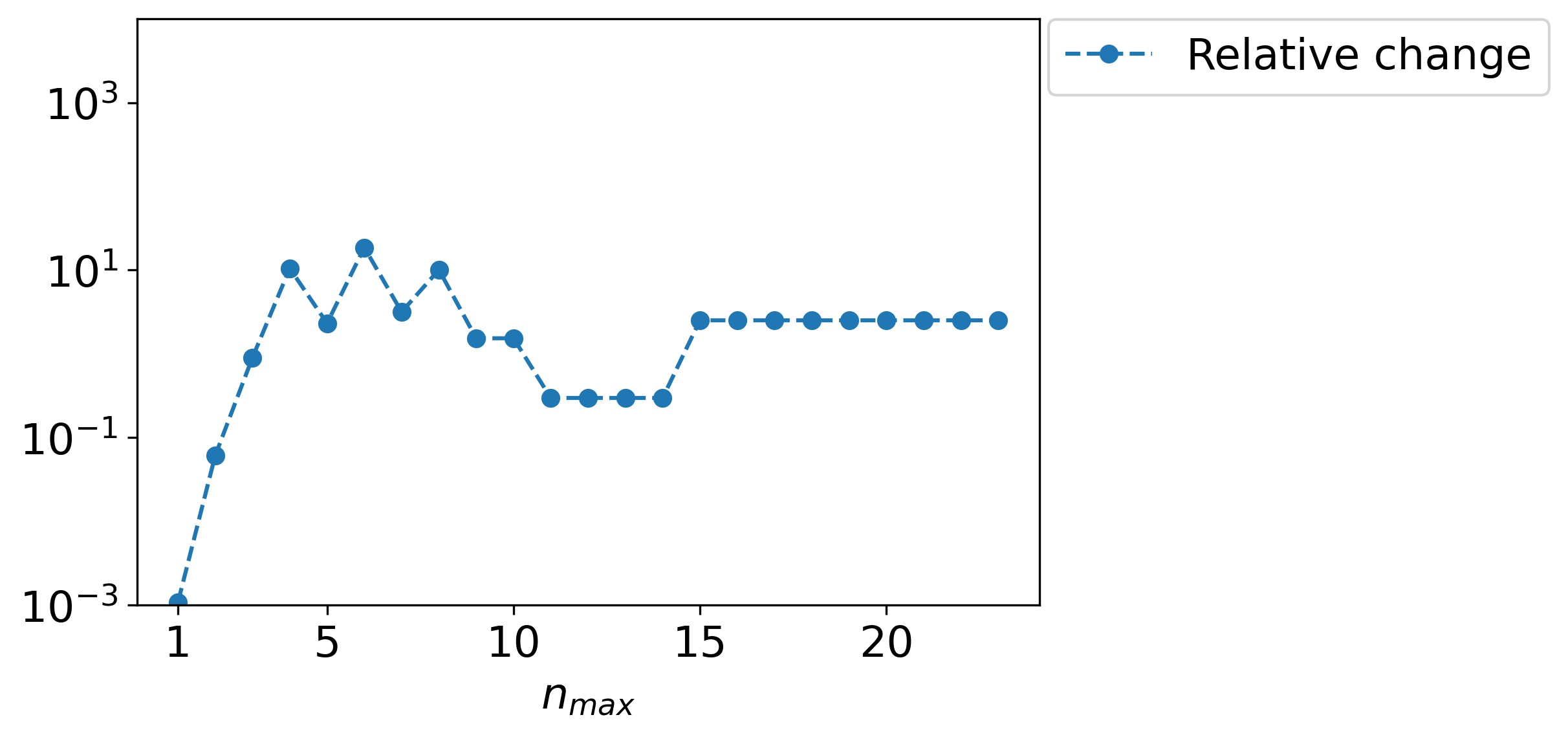}
\includegraphics[width=0.49\linewidth]{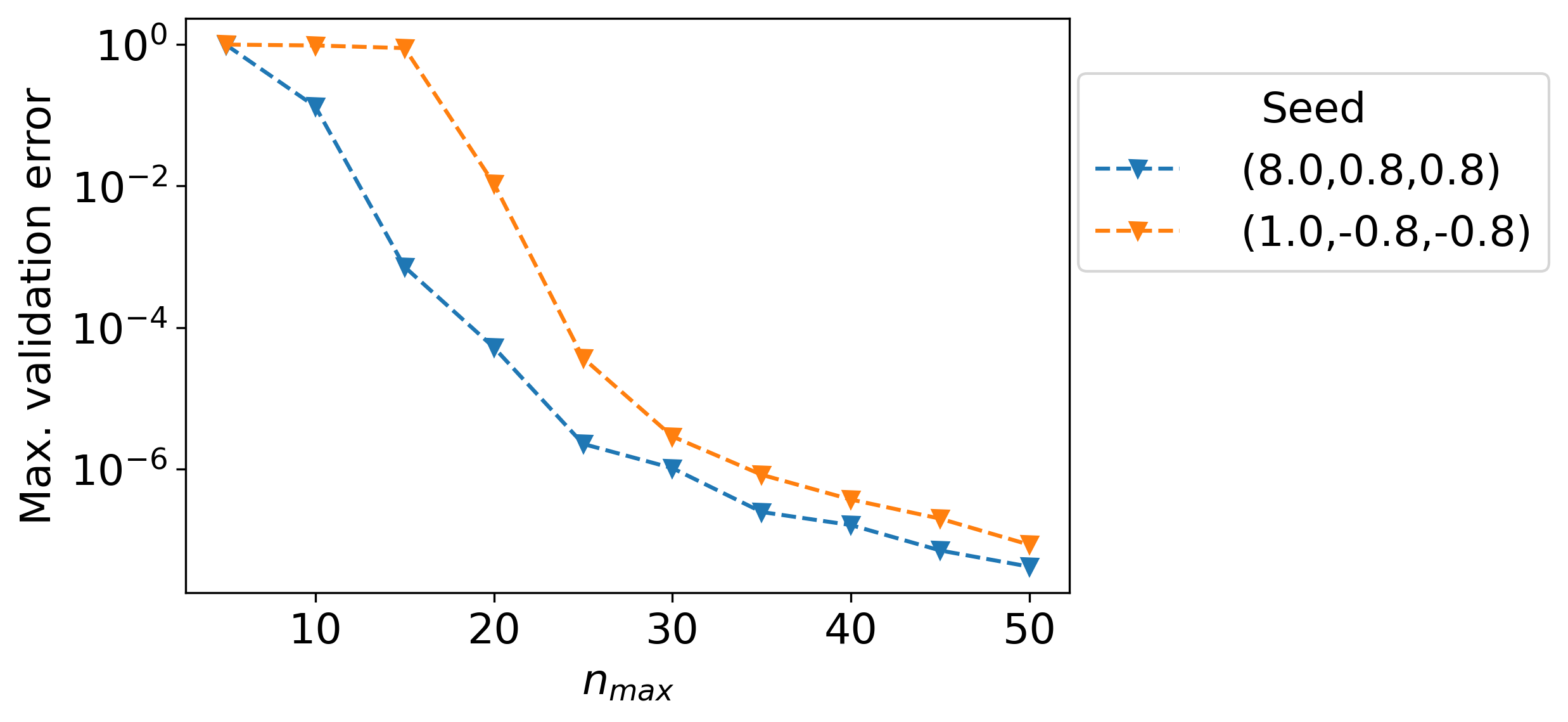}   
\includegraphics[width=0.49\linewidth]{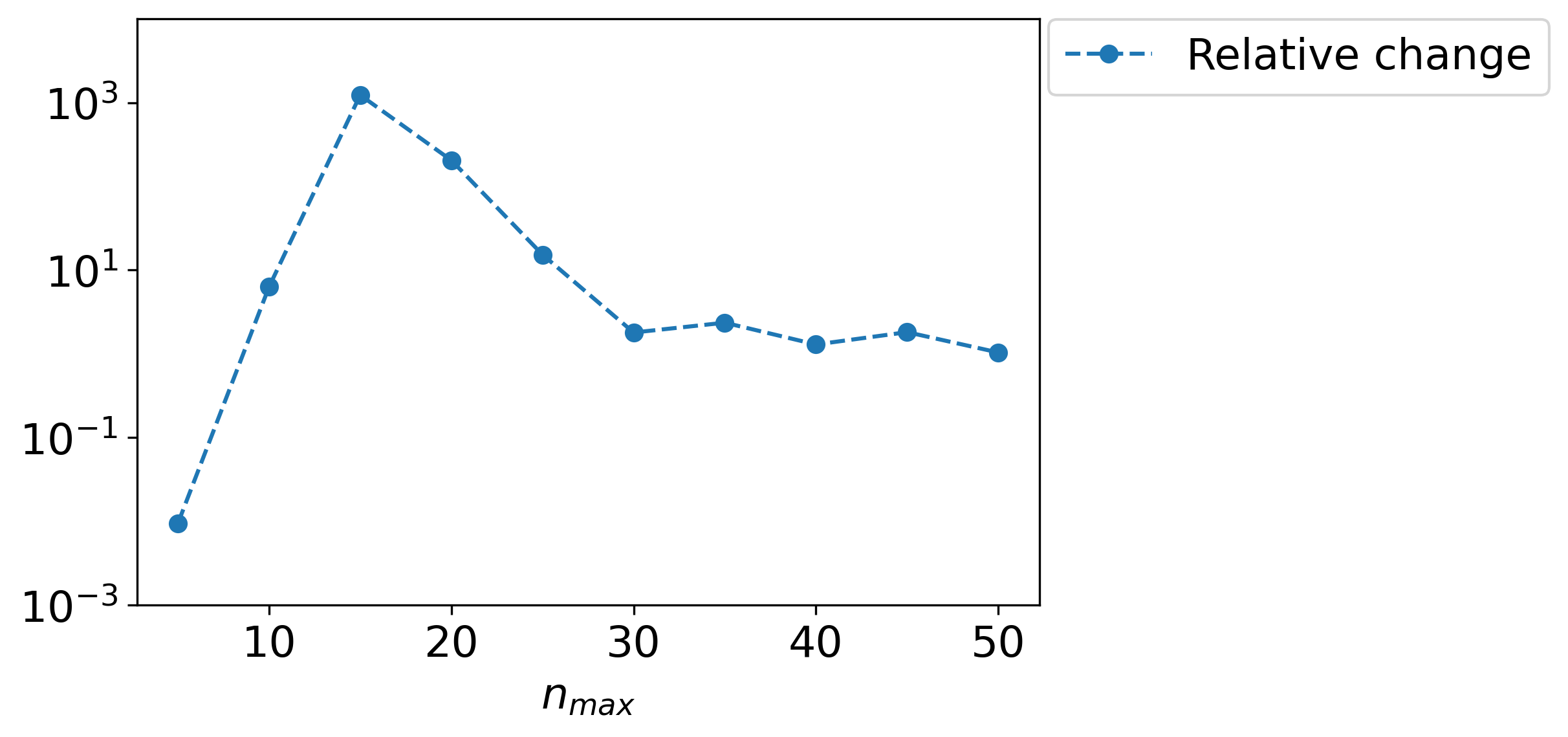}
\includegraphics[width=0.49\linewidth]{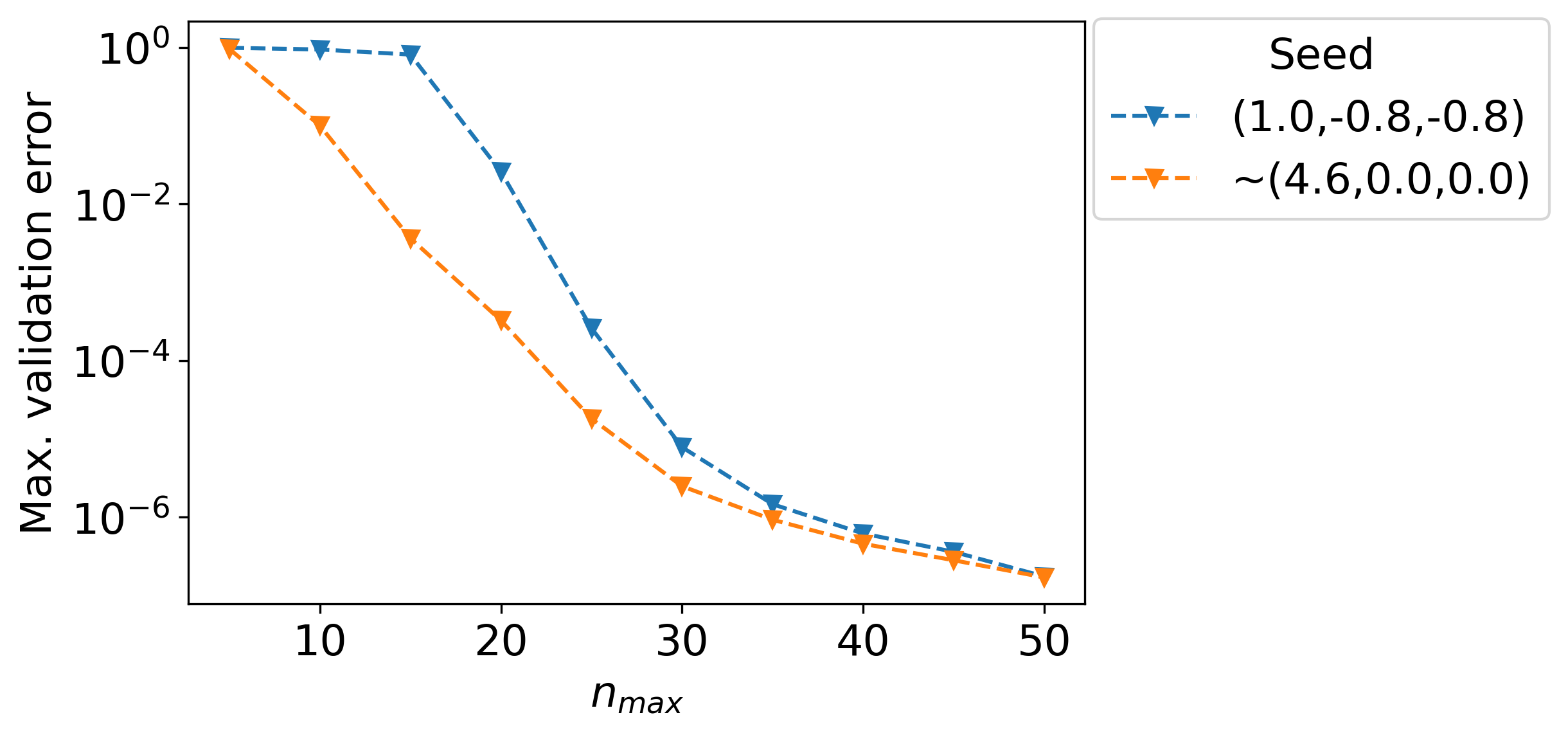}   
\includegraphics[width=0.49\linewidth]{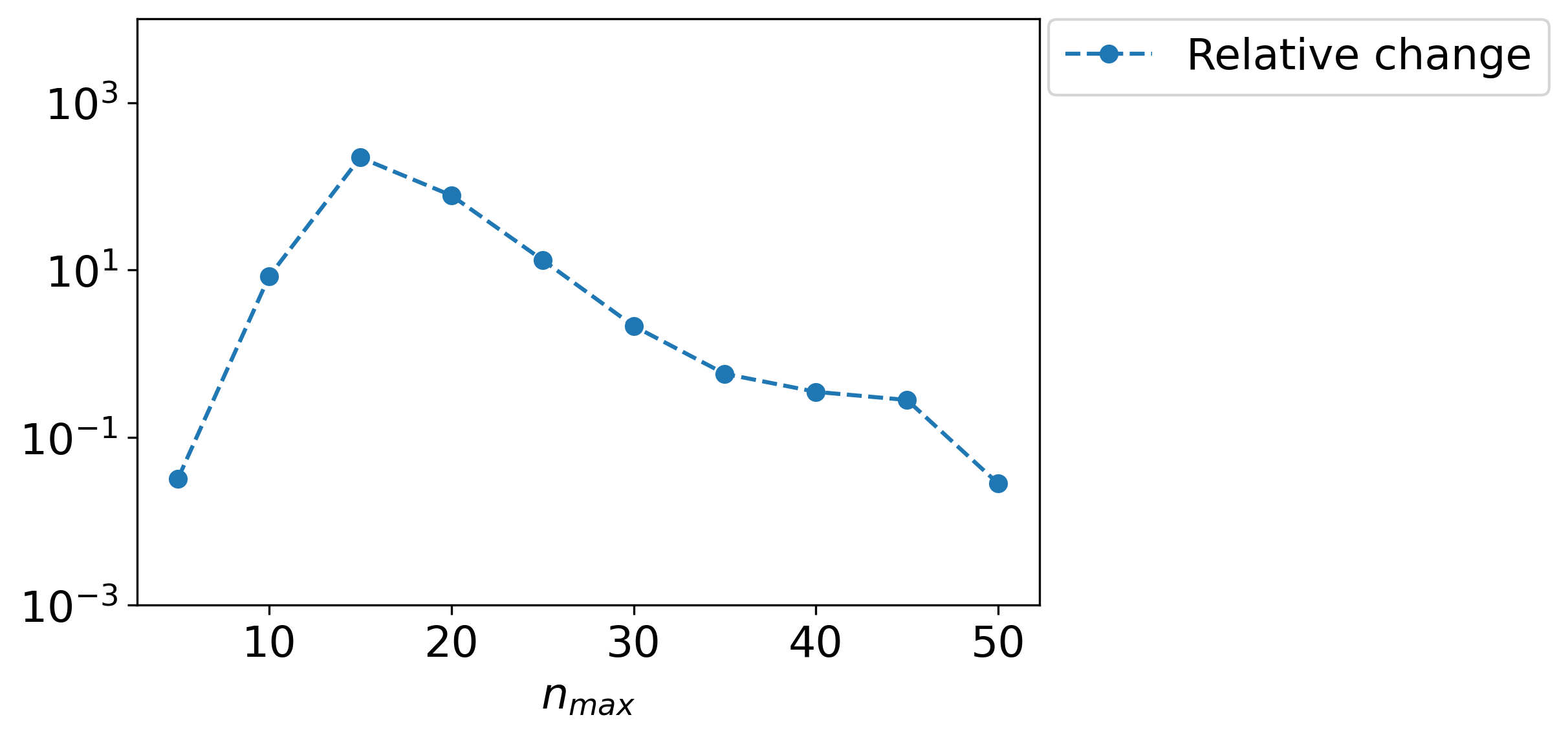}
    \caption{From top to bottom, hp-greedy results are shown for 1D, 2D and 3D cases, using different seeds. Each curve in the left panels shows the highest accuracy model for each $n_{max}$, choosing the ``optimal'' value of ${\ell}_{max}$.  Unlike a global, standard reduced basis approach, in hp-greedy there is a strong dependence of the achieved errors on the seeds used. The right panels show the ratios between the errors of the two models of each left panel; we can observe up to three orders of magnitude of difference in the errors when varying the seed.}
    \label{fig:seeds}
\end{figure}

\begin{figure}[ht]
\centering
\hspace{6mm}               
 \includegraphics[width=0.425\linewidth]{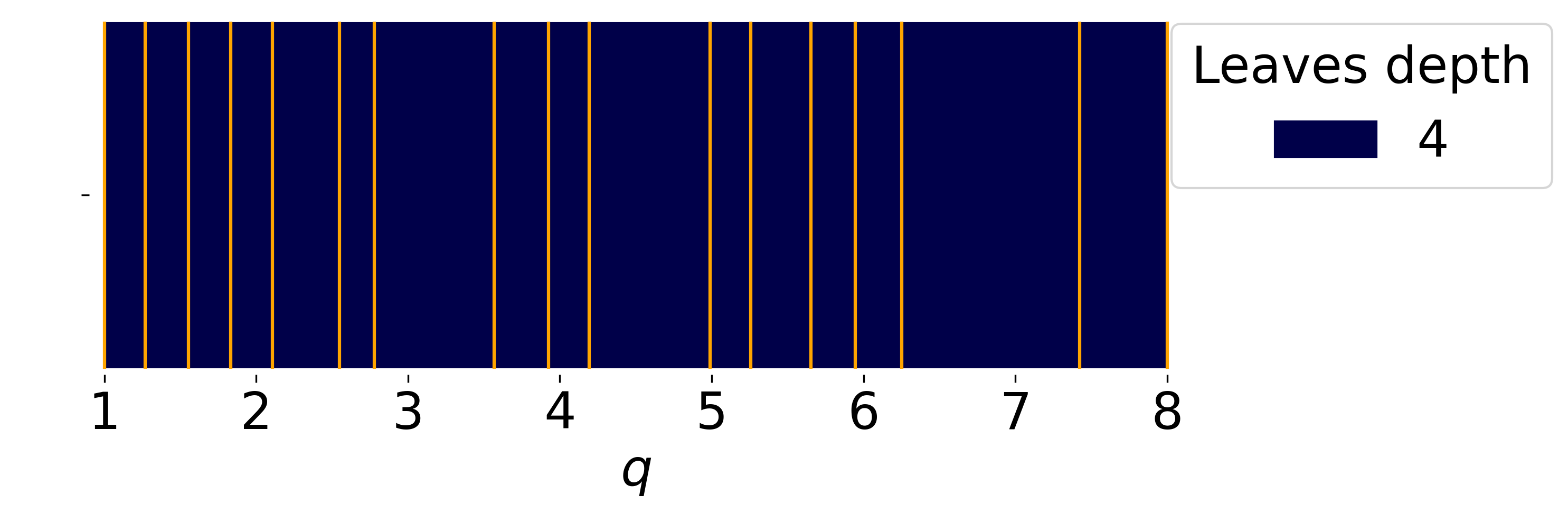}
\hspace{8mm}                
 \includegraphics[width=0.425\linewidth]{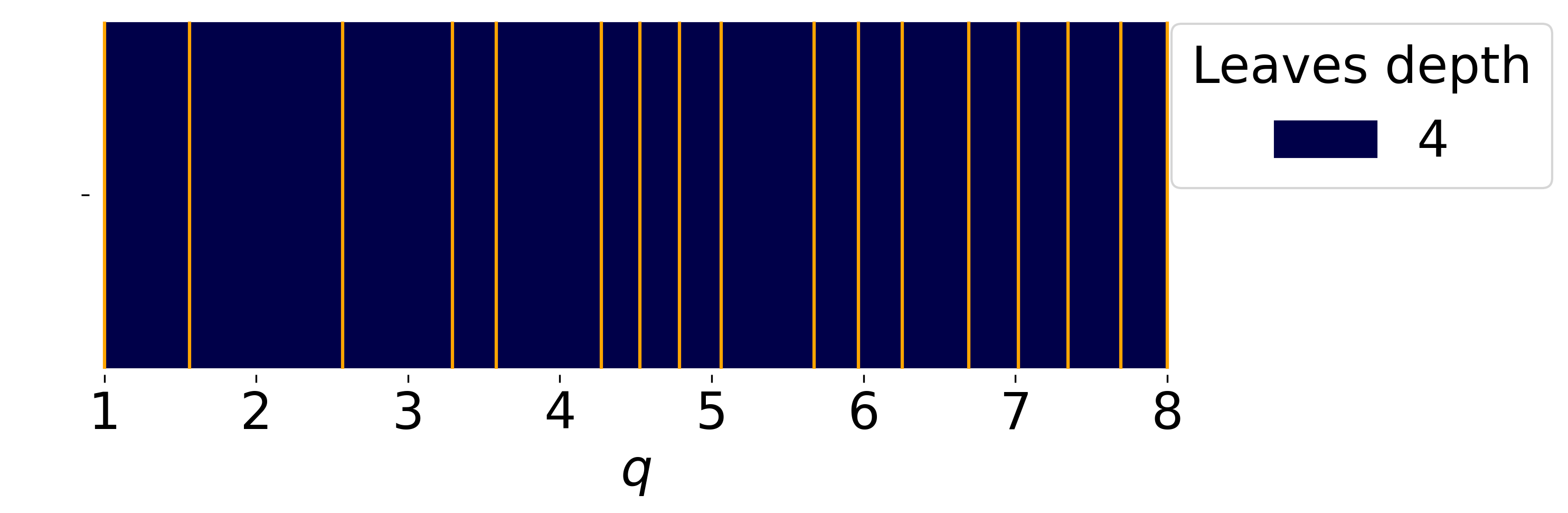}
    \caption{Domain decompositions obtained with hp-greedy in the 1D gravitational wave case, using mass ratios $q=1$ and $q=8$ as seeds (left and right panel, respectively), $n_{max}=5$ and $l_{max}=4$. The dependence on the seed can be qualitatively noticed. }
    \label{fig:gw_1d_part}
\end{figure}

\subsection{Convergence} \label{sec:c_o}

We assessed the convergence of hp-greedy with respect to $n_{max}$, choosing for each value the highest accuracy model when varying the seed and ${\ell}_{max}$, and comparing against a global basis (i.e., the standard approach). The results for the 1D, 2D, and 3D cases are shown in Figure~\ref{fig:convergence_gw}. On one hand, there are always accuracy improvements at fixed $n_{max}$, in some cases by several orders of magnitude.
On the other hand, for all cases, bases with lower dimension $n_{max}$ are obtained for a fixed accuracy. As we discuss in Section~\ref{sec:com}, smaller values of $n_{max}$ are related to shorter evaluation times and hence faster statistical inference. 
For hp-greedy, we notice an exponential convergence from the onset, $n_{max} \geq 1$. In contrast, for a global basis, this spectral convergence appears asymptotically, i.e., for sufficiently large values of $n_{max}$. 

\begin{figure}[ht]
    \centering
    \includegraphics[width=0.47\linewidth]{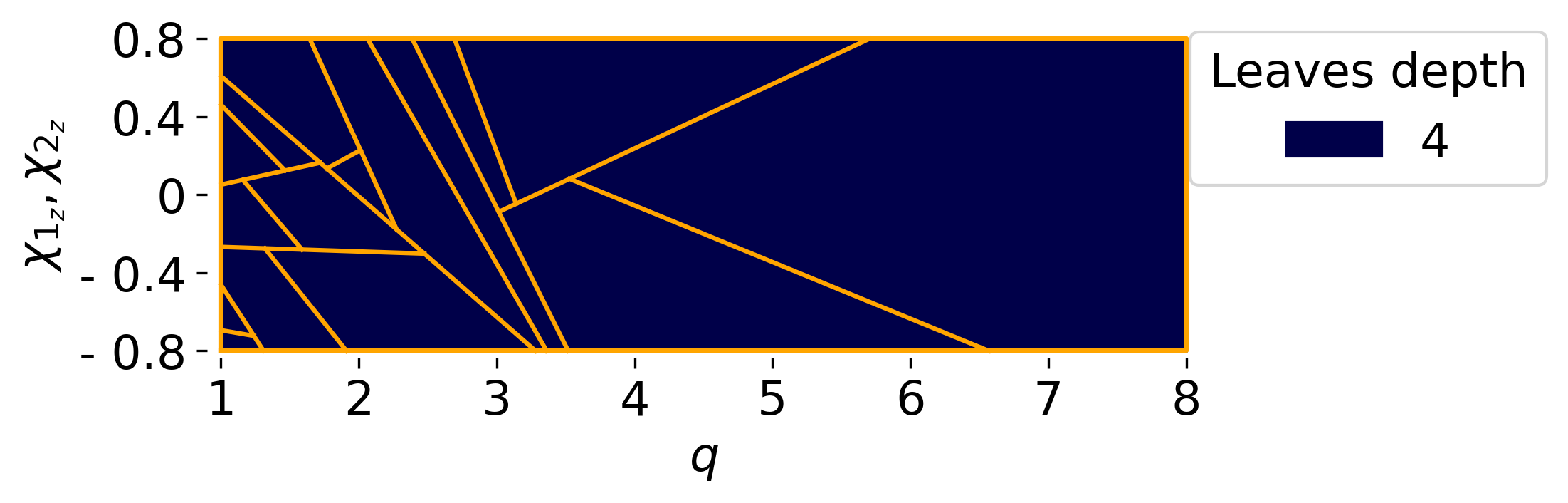}
    \includegraphics[width=0.47\linewidth]{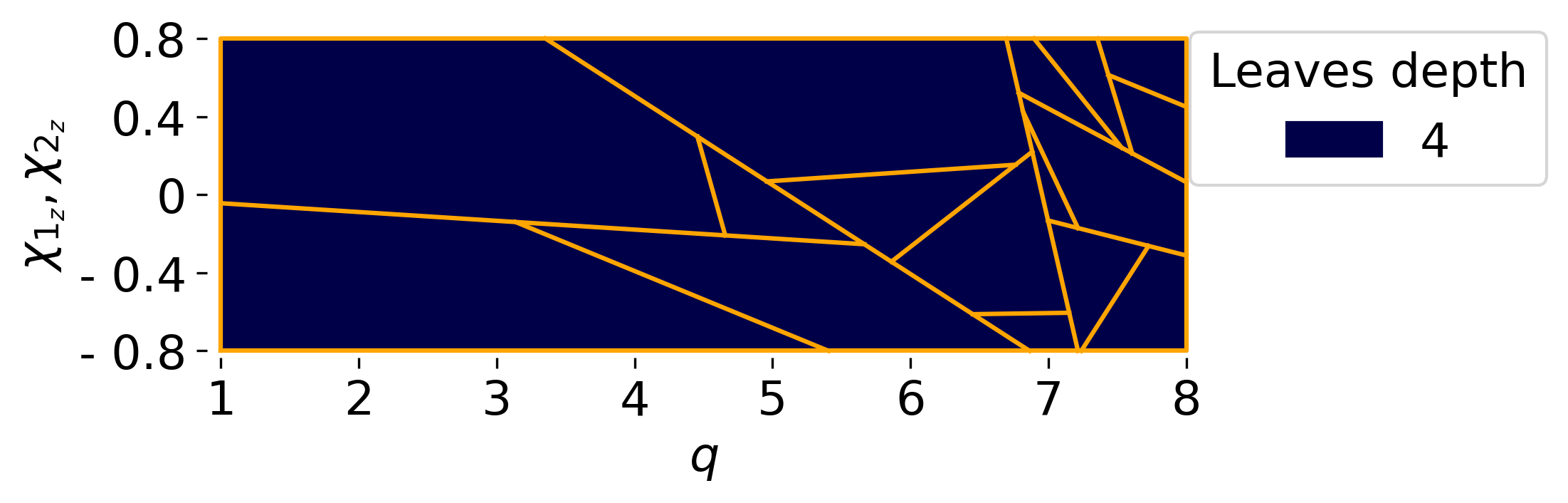}
    \caption{Domain decompositions obtained with hp-greedy using different seeds for the 2D case. In order, from left to right, the figures have seeds located at $(1,-0.8,-0.8)$ and $(8,0.8,0.8)$. For pure visualization purposes and definiteness, hyperparameter values ${\ell}_{max}=4$ and $n_{max}=20$ were arbitrarily used. In these cases the leaves have the same depth, and also the highest possible. This means that lower depths were not enough for every subspace and the algorithm stopped in each leaf because of the value of ${\ell}_{max}$. The impact of the seed choice on the domain decomposition is apparent.  }
    \label{fig:gw_2d_part}
\end{figure}

 \begin{figure}[ht]
    \centering 
   \includegraphics[width=0.49\linewidth]{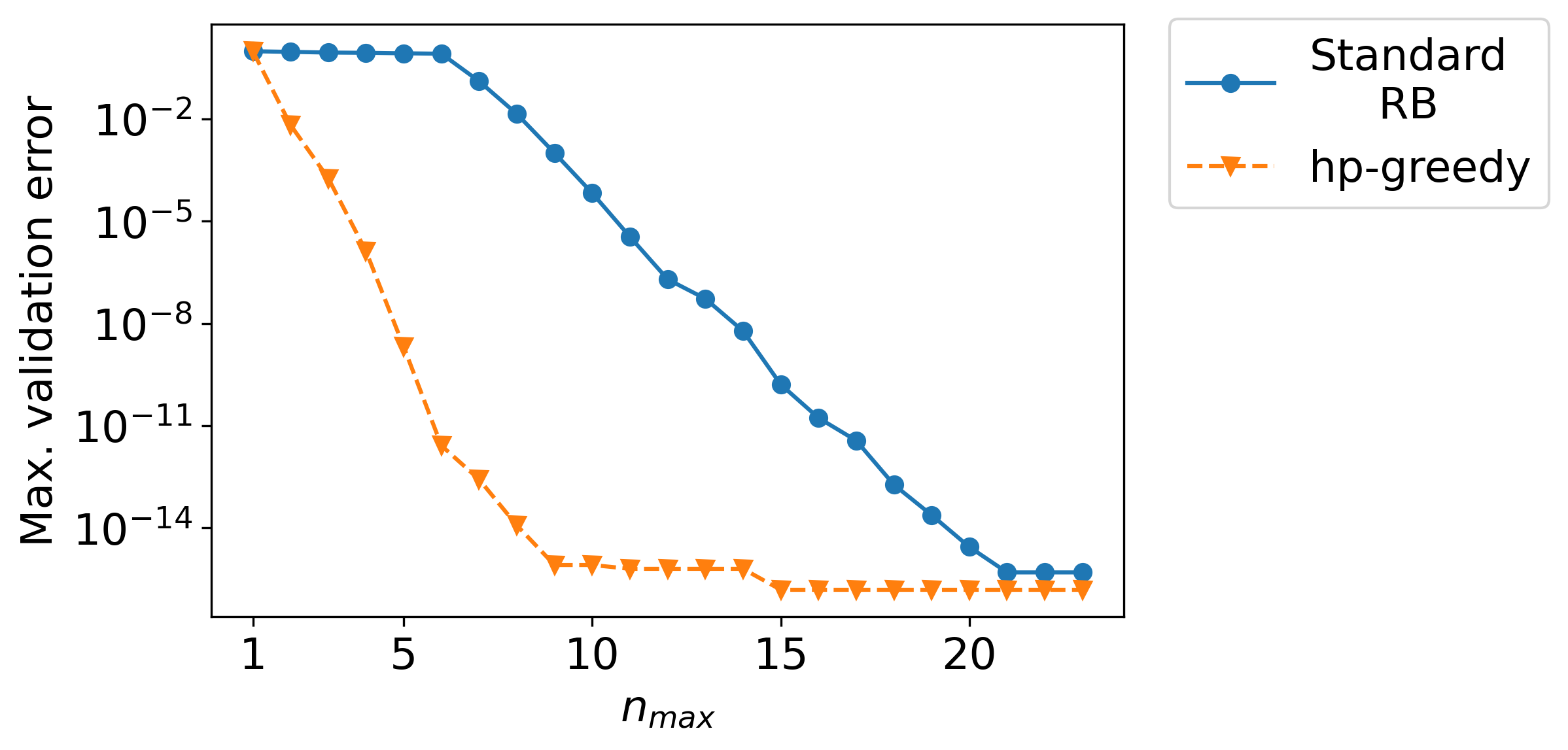}
   \includegraphics[width=0.49\linewidth]{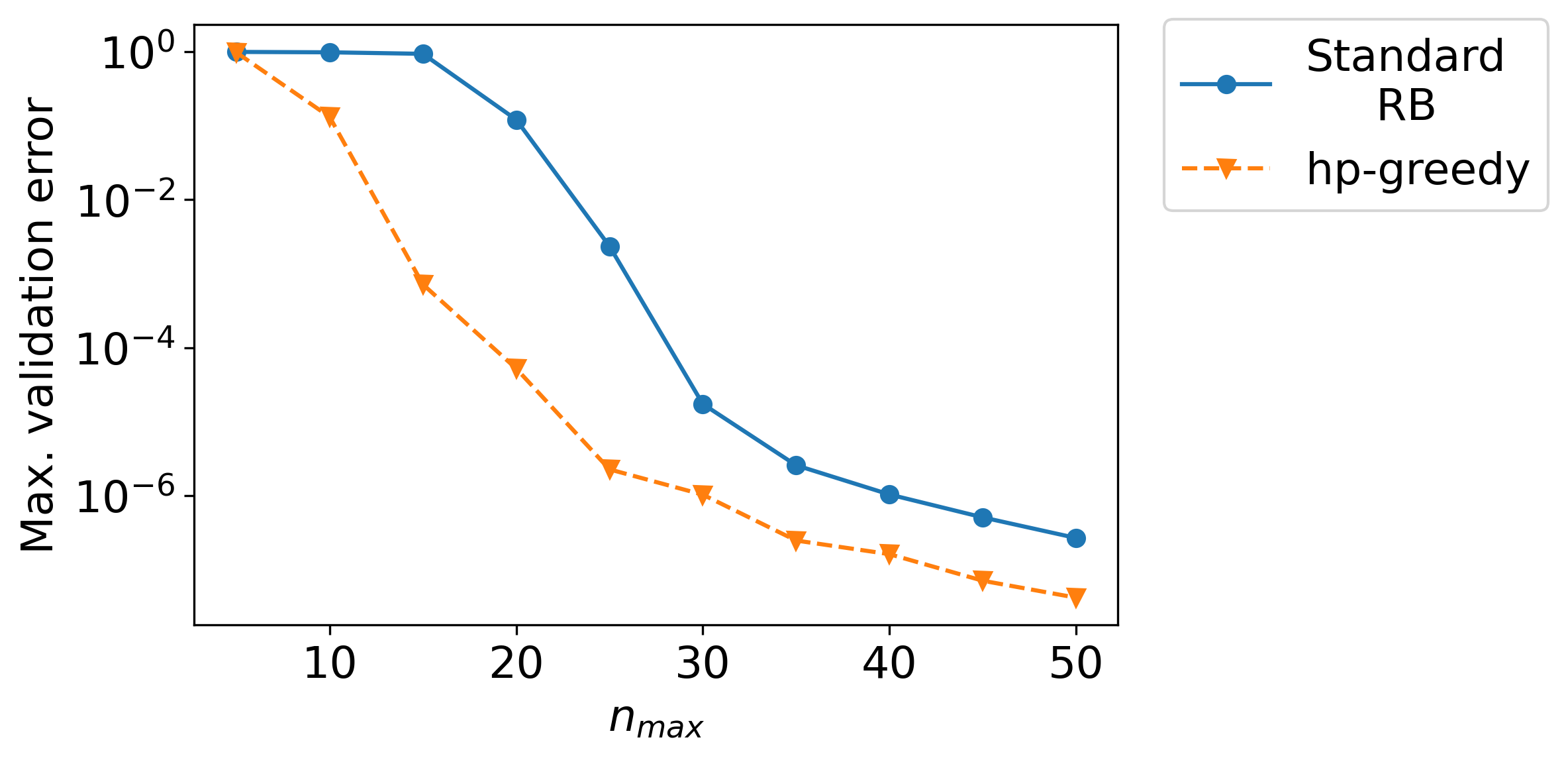}
   \includegraphics[width=0.49\linewidth]{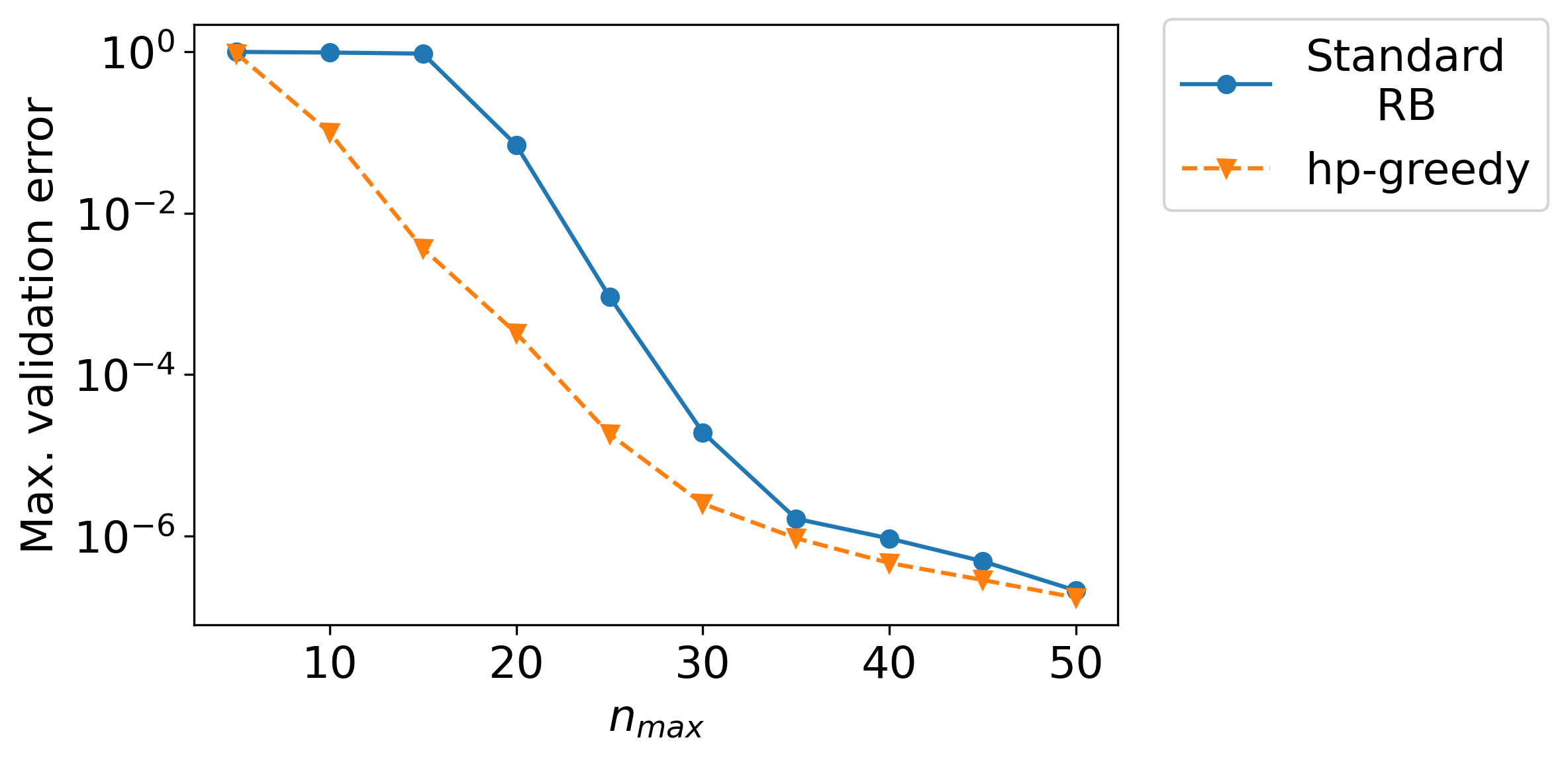}    
    \caption{Convergence of best performing hp-greedy models compared to the one of a global reduced basis (labeled as ``Standard RB'') for the 1D, 2D and 3D cases, from left to right and top to bottom.   }
    \label{fig:convergence_gw}
\end{figure}

\section{Discussion and Future work} \label{sec:com}

In this work, the usage of hp-greedy refinement in reduced basis within GW science is proposed, as a framework for partitioning the parameter space in an unstructured way and building a set of local reduced basis. This framework is a generalization of the standard reduced basis approach (which can be obtained as a special case by setting the hyperparameter $l_{max}=0$), and aims at improving the accuracy and compactness of a reduced basis representation. 

One main conclusion from our numerical experiments with hp-greedy is that the seed of the algorithm should be treated as an hyperparameter, in addition to $n_{max}$ and $\ell_{max}$, since in general it does affect the accuracy of the resulting reduced bases.
This behavior differs from the standard reduced basis framework, in which there is no partitioning of the parameter space, and the accuracy of the basis is insensitive to the seed choice. In addition, in the applications studied here, the seed choice does show a qualitatively noticeable impact on the structure of the domain partition.

It is known that a global reduced basis obtained with a greedy approach is a quasi-optimal approximation. This means that it is difficult (or theoretically almost impossible) to reduce its dimensionality without accuracy loss: hp-greedy overcomes this challenge by partitioning the parameter space. 
From a simple two-dimensional model with a discontinuity in the parameter space, we concluded that the approach works with a reduction of the dimensionality needed for the bases to represent the function space for almost every precision used.
Moreover, we have found that this improvement holds when applied to the more realistic case of two colliding spinning black holes, up to three dimensions, which is the highest dimensionality considered in this work. We have also found improvements in the accuracy for a fixed maximum dimensionality of the bases, $n_{max}$.

We envision several applications of our approach. One potential use case is to further accelerate statistical inference, for example in parameter estimation of the source of a detected gravitational wave, using reduced order quadratures (ROQ)~\cite{antil2012two,Canizares:2013ywa,Morisaki_2020,PhysRevLett.114.071104,PhysRevD.94.044031}. In more detail, parameter estimation serves to compute likelihoods, which involve integrals being often computationally expensive and requiring multiple on-demand sequential evaluations (e.g., via Markov chain Monte Carlo simulations). ROQ accelerates this process by constructing an application-specific quadrature rule using a reduced basis and the Empirical Interpolation Method~\cite{Canizares:2013ywa}, with a cost proportional to the dimensionality of the basis. Along this line, ROQ likelihood computations can be accelerated by using a multi-domain hp-greedy basis, since, for any given precision, the dimensionality of each local basis is expected to be smaller than that one of a global approach. This perspective would be similar in spirit to focused ROQ~\cite{Morisaki_2020}, where a reduced basis is constructed in a region close to the parameters found in the trigger part of the detection pipeline.

In any application of hp-greedy, the domain partition corresponding to any value of a parameter $\lambda$ must be found in order to use the sub-space associated with a leaf that contains that value. Having a tree structure to divide the parameter space allows us to perform a fast search. This entails a binary search in which each node contains a pair of anchor points $\hat{\Lambda}_{V_1}$ and $\hat{\Lambda}_{V_2}$. Then, for a given parameter value $\lambda$, we can compute its distance to a pair of anchor points to descend one level in the tree, choosing the node that contains the closest anchor point with respect to that distance. The distance comparisons can begin with the anchor points of the root of the tree until reaching a leaf, being associated with a local reduced basis, that represents the sub-space where $\lambda$ is located. For example, if the tree is balanced and there are $n$ sub-spaces as leaves, the computational cost of the search becomes $\mathcal{O}(\log n)$, which improves the $\mathcal{O}(n)$ cost required if one looks at each subspace one by one. 

Besides surrogate modeling, a second use case for hp-greedy is the search for gravitational waves using a nearest neighbors strategy, instead of the standard direct approach of comparing each candidate signal with a bank of templates one by one. 

Another natural application can be one in which a physical discontinuity in the parameter space is present, and thus, a global reduced basis is likely to show slow convergence due to Gibb's phenomena~\cite{hesthaven2007spectral}. Some examples might include: i) two compact objects with a non-vanishing impact parameter, for which there can be fly-off, or collision~\cite{Pretorius_2007}, ii) the merger of two neutron stars or a mixed pair of a black hole and a neutron star, for which there can be a merger into a larger neutron star, or to a black hole~\cite{Shibata_NR}. 

In future work, we plan to perform a systematic hyperparameter optimization to devise rules for choosing appropriate seeds, $n_{max}$ and $\ell_{max}$, allowing us to build hp-greedy models with high accuracy in a faster way.

\section{Acknowledgments} \label{sec:ack}
This work was partially supported by CONICET-Argentina. We thank Marcelo Rubio and Atuel Villegas for their feedback on a previous version of this manuscript.  

\bibliography{references}

\begin{thebibliography}{43}%
\makeatletter
\providecommand \@ifxundefined [1]{%
 \@ifx{#1\undefined}
}%
\providecommand \@ifnum [1]{%
 \ifnum #1\expandafter \@firstoftwo
 \else \expandafter \@secondoftwo
 \fi
}%
\providecommand \@ifx [1]{%
 \ifx #1\expandafter \@firstoftwo
 \else \expandafter \@secondoftwo
 \fi
}%
\providecommand \natexlab [1]{#1}%
\providecommand \enquote  [1]{``#1''}%
\providecommand \bibnamefont  [1]{#1}%
\providecommand \bibfnamefont [1]{#1}%
\providecommand \citenamefont [1]{#1}%
\providecommand \href@noop [0]{\@secondoftwo}%
\providecommand \href [0]{\begingroup \@sanitize@url \@href}%
\providecommand \@href[1]{\@@startlink{#1}\@@href}%
\providecommand \@@href[1]{\endgroup#1\@@endlink}%
\providecommand \@sanitize@url [0]{\catcode `\\12\catcode `\$12\catcode
  `\&12\catcode `\#12\catcode `\^12\catcode `\_12\catcode `\%12\relax}%
\providecommand \@@startlink[1]{}%
\providecommand \@@endlink[0]{}%
\providecommand \url  [0]{\begingroup\@sanitize@url \@url }%
\providecommand \@url [1]{\endgroup\@href {#1}{\urlprefix }}%
\providecommand \urlprefix  [0]{URL }%
\providecommand \Eprint [0]{\href }%
\providecommand \doibase [0]{https://doi.org/}%
\providecommand \selectlanguage [0]{\@gobble}%
\providecommand \bibinfo  [0]{\@secondoftwo}%
\providecommand \bibfield  [0]{\@secondoftwo}%
\providecommand \translation [1]{[#1]}%
\providecommand \BibitemOpen [0]{}%
\providecommand \bibitemStop [0]{}%
\providecommand \bibitemNoStop [0]{.\EOS\space}%
\providecommand \EOS [0]{\spacefactor3000\relax}%
\providecommand \BibitemShut  [1]{\csname bibitem#1\endcsname}%
\let\auto@bib@innerbib\@empty
\bibitem [{Abb(2016)}]{Abbott_2016}%
  \BibitemOpen
  \bibfield  {title} {\bibinfo {title} {Observation of gravitational waves from
  a binary black hole merger},\ }\href
  {https://doi.org/10.1103/PhysRevLett.116.061102} {\bibfield  {journal}
  {\bibinfo  {journal} {Phys. Rev. Lett.}\ }\textbf {\bibinfo {volume} {116}},\
  \bibinfo {pages} {061102} (\bibinfo {year} {2016})}\BibitemShut {NoStop}%
\bibitem [{\citenamefont {Lehner}\ and\ \citenamefont
  {Pretorius}(2014)}]{Lehner_2014}%
  \BibitemOpen
  \bibfield  {author} {\bibinfo {author} {\bibfnamefont {L.}~\bibnamefont
  {Lehner}}\ and\ \bibinfo {author} {\bibfnamefont {F.}~\bibnamefont
  {Pretorius}},\ }\bibfield  {title} {\bibinfo {title} {Numerical relativity
  and astrophysics},\ }\href
  {https://doi.org/10.1146/annurev-astro-081913-040031} {\bibfield  {journal}
  {\bibinfo  {journal} {Annual Review of Astronomy and Astrophysics}\ }\textbf
  {\bibinfo {volume} {52}},\ \bibinfo {pages} {661} (\bibinfo {year}
  {2014})}\BibitemShut {NoStop}%
\bibitem [{\citenamefont {Veitch}\ \emph {et~al.}(2015)\citenamefont {Veitch},
  \citenamefont {Raymond}, \citenamefont {Farr}, \citenamefont {Farr},
  \citenamefont {Graff}, \citenamefont {Vitale}, \citenamefont {Aylott},
  \citenamefont {Blackburn}, \citenamefont {Christensen}, \citenamefont
  {Coughlin}, \citenamefont {Pozzo}, \citenamefont {Feroz}, \citenamefont
  {Gair}, \citenamefont {Haster}, \citenamefont {Kalogera}, \citenamefont
  {Littenberg}, \citenamefont {Mandel}, \citenamefont {O'Shaughnessy},
  \citenamefont {Pitkin}, \citenamefont {Rodriguez}, \citenamefont {Röver},
  \citenamefont {Sidery}, \citenamefont {Smith}, \citenamefont {Sluys},
  \citenamefont {Vecchio}, \citenamefont {Vousden},\ and\ \citenamefont
  {Wade}}]{Veitch_2015}%
  \BibitemOpen
  \bibfield  {author} {\bibinfo {author} {\bibfnamefont {J.}~\bibnamefont
  {Veitch}}, \bibinfo {author} {\bibfnamefont {V.}~\bibnamefont {Raymond}},
  \bibinfo {author} {\bibfnamefont {B.}~\bibnamefont {Farr}}, \bibinfo {author}
  {\bibfnamefont {W.}~\bibnamefont {Farr}}, \bibinfo {author} {\bibfnamefont
  {P.}~\bibnamefont {Graff}}, \bibinfo {author} {\bibfnamefont
  {S.}~\bibnamefont {Vitale}}, \bibinfo {author} {\bibfnamefont
  {B.}~\bibnamefont {Aylott}}, \bibinfo {author} {\bibfnamefont
  {K.}~\bibnamefont {Blackburn}}, \bibinfo {author} {\bibfnamefont
  {N.}~\bibnamefont {Christensen}}, \bibinfo {author} {\bibfnamefont
  {M.}~\bibnamefont {Coughlin}}, \bibinfo {author} {\bibfnamefont {W.~D.}\
  \bibnamefont {Pozzo}}, \bibinfo {author} {\bibfnamefont {F.}~\bibnamefont
  {Feroz}}, \bibinfo {author} {\bibfnamefont {J.}~\bibnamefont {Gair}},
  \bibinfo {author} {\bibfnamefont {C.-J.}\ \bibnamefont {Haster}}, \bibinfo
  {author} {\bibfnamefont {V.}~\bibnamefont {Kalogera}}, \bibinfo {author}
  {\bibfnamefont {T.}~\bibnamefont {Littenberg}}, \bibinfo {author}
  {\bibfnamefont {I.}~\bibnamefont {Mandel}}, \bibinfo {author} {\bibfnamefont
  {R.}~\bibnamefont {O'Shaughnessy}}, \bibinfo {author} {\bibfnamefont
  {M.}~\bibnamefont {Pitkin}}, \bibinfo {author} {\bibfnamefont
  {C.}~\bibnamefont {Rodriguez}}, \bibinfo {author} {\bibfnamefont
  {C.}~\bibnamefont {Röver}}, \bibinfo {author} {\bibfnamefont
  {T.}~\bibnamefont {Sidery}}, \bibinfo {author} {\bibfnamefont
  {R.}~\bibnamefont {Smith}}, \bibinfo {author} {\bibfnamefont {M.~V.~D.}\
  \bibnamefont {Sluys}}, \bibinfo {author} {\bibfnamefont {A.}~\bibnamefont
  {Vecchio}}, \bibinfo {author} {\bibfnamefont {W.}~\bibnamefont {Vousden}},\
  and\ \bibinfo {author} {\bibfnamefont {L.}~\bibnamefont {Wade}},\ }\bibfield
  {title} {\bibinfo {title} {Parameter estimation for compact binaries with
  ground-based gravitational-wave observations using the {LALInference}
  software library},\ }\href {https://doi.org/10.1103%2Fphysrevd.91.042003}
  {\bibfield  {journal} {\bibinfo  {journal} {Physical Review D}\ }\textbf
  {\bibinfo {volume} {91}} (\bibinfo {year} {2015})}\BibitemShut {NoStop}%
\bibitem [{\citenamefont {Christensen}\ and\ \citenamefont
  {Meyer}(2022)}]{Christensen_2022}%
  \BibitemOpen
  \bibfield  {author} {\bibinfo {author} {\bibfnamefont {N.}~\bibnamefont
  {Christensen}}\ and\ \bibinfo {author} {\bibfnamefont {R.}~\bibnamefont
  {Meyer}},\ }\bibfield  {title} {\bibinfo {title} {Parameter estimation with
  gravitational waves},\ }\href
  {https://doi.org/10.1103%2Frevmodphys.94.025001} {\bibfield  {journal}
  {\bibinfo  {journal} {Reviews of Modern Physics}\ }\textbf {\bibinfo {volume}
  {94}} (\bibinfo {year} {2022})}\BibitemShut {NoStop}%
\bibitem [{\citenamefont {Canizares}\ \emph {et~al.}(2013)\citenamefont
  {Canizares}, \citenamefont {Field}, \citenamefont {Gair},\ and\ \citenamefont
  {Tiglio}}]{Canizares:2013ywa}%
  \BibitemOpen
  \bibfield  {author} {\bibinfo {author} {\bibfnamefont {P.}~\bibnamefont
  {Canizares}}, \bibinfo {author} {\bibfnamefont {S.~E.}\ \bibnamefont
  {Field}}, \bibinfo {author} {\bibfnamefont {J.~R.}\ \bibnamefont {Gair}},\
  and\ \bibinfo {author} {\bibfnamefont {M.}~\bibnamefont {Tiglio}},\
  }\bibfield  {title} {\bibinfo {title} {{Gravitational wave parameter
  estimation with compressed likelihood evaluations}},\ }\href
  {https://doi.org/10.1103/PhysRevD.87.124005} {\bibfield  {journal} {\bibinfo
  {journal} {Phys. Rev.}\ }\textbf {\bibinfo {volume} {D87}},\ \bibinfo {pages}
  {124005} (\bibinfo {year} {2013})}\BibitemShut {NoStop}%
\bibitem [{\citenamefont {Owen}\ and\ \citenamefont
  {Sathyaprakash}(1999)}]{Owen:1998dk}%
  \BibitemOpen
  \bibfield  {author} {\bibinfo {author} {\bibfnamefont {B.~J.}\ \bibnamefont
  {Owen}}\ and\ \bibinfo {author} {\bibfnamefont {B.}~\bibnamefont
  {Sathyaprakash}},\ }\bibfield  {title} {\bibinfo {title} {{Matched filtering
  of gravitational waves from inspiraling compact binaries: Computational cost
  and template placement}},\ }\href
  {https://doi.org/10.1103/PhysRevD.60.022002} {\bibfield  {journal} {\bibinfo
  {journal} {Phys. Rev.}\ }\textbf {\bibinfo {volume} {D60}},\ \bibinfo {pages}
  {022002} (\bibinfo {year} {1999})}\BibitemShut {NoStop}%
\bibitem [{\citenamefont {Cannon}\ \emph {et~al.}(2012)\citenamefont {Cannon},
  \citenamefont {Cariou}, \citenamefont {Chapman}, \citenamefont
  {Crispin-Ortuzar}, \citenamefont {Fotopoulos}, \citenamefont {Frei},
  \citenamefont {Hanna}, \citenamefont {Kara}, \citenamefont {Keppel},
  \citenamefont {Liao}, \citenamefont {Privitera}, \citenamefont {Searle},
  \citenamefont {Singer},\ and\ \citenamefont {Weinstein}}]{Cannon_2012}%
  \BibitemOpen
  \bibfield  {author} {\bibinfo {author} {\bibfnamefont {K.}~\bibnamefont
  {Cannon}}, \bibinfo {author} {\bibfnamefont {R.}~\bibnamefont {Cariou}},
  \bibinfo {author} {\bibfnamefont {A.}~\bibnamefont {Chapman}}, \bibinfo
  {author} {\bibfnamefont {M.}~\bibnamefont {Crispin-Ortuzar}}, \bibinfo
  {author} {\bibfnamefont {N.}~\bibnamefont {Fotopoulos}}, \bibinfo {author}
  {\bibfnamefont {M.}~\bibnamefont {Frei}}, \bibinfo {author} {\bibfnamefont
  {C.}~\bibnamefont {Hanna}}, \bibinfo {author} {\bibfnamefont
  {E.}~\bibnamefont {Kara}}, \bibinfo {author} {\bibfnamefont {D.}~\bibnamefont
  {Keppel}}, \bibinfo {author} {\bibfnamefont {L.}~\bibnamefont {Liao}},
  \bibinfo {author} {\bibfnamefont {S.}~\bibnamefont {Privitera}}, \bibinfo
  {author} {\bibfnamefont {A.}~\bibnamefont {Searle}}, \bibinfo {author}
  {\bibfnamefont {L.}~\bibnamefont {Singer}},\ and\ \bibinfo {author}
  {\bibfnamefont {A.}~\bibnamefont {Weinstein}},\ }\bibfield  {title} {\bibinfo
  {title} {Toward early-warning detection of gravitational waves from compact
  binary coalescence},\ }\href {https://doi.org/10.1088/0004-637X/748/2/136}
  {\bibfield  {journal} {\bibinfo  {journal} {The Astrophysical Journal}\
  }\textbf {\bibinfo {volume} {748}},\ \bibinfo {pages} {136} (\bibinfo {year}
  {2012})}\BibitemShut {NoStop}%
\bibitem [{\citenamefont {Babak}\ \emph {et~al.}(2013)\citenamefont {Babak},
  \citenamefont {Biswas}, \citenamefont {Brady}, \citenamefont {Brown},
  \citenamefont {Cannon}, \citenamefont {Capano}, \citenamefont {Clayton},
  \citenamefont {Cokelaer}, \citenamefont {Creighton}, \citenamefont {Dent},
  \citenamefont {Dietz}, \citenamefont {Fairhurst}, \citenamefont {Fotopoulos},
  \citenamefont {Gonz{\'{a} }lez}, \citenamefont {Hanna}, \citenamefont
  {Harry}, \citenamefont {Jones}, \citenamefont {Keppel}, \citenamefont
  {McKechan}, \citenamefont {Pekowsky}, \citenamefont {Privitera},
  \citenamefont {Robinson}, \citenamefont {Rodriguez}, \citenamefont
  {Sathyaprakash}, \citenamefont {Sengupta}, \citenamefont {Vallisneri},
  \citenamefont {Vaulin},\ and\ \citenamefont {Weinstein}}]{Babak_2013}%
  \BibitemOpen
  \bibfield  {author} {\bibinfo {author} {\bibfnamefont {S.}~\bibnamefont
  {Babak}}, \bibinfo {author} {\bibfnamefont {R.}~\bibnamefont {Biswas}},
  \bibinfo {author} {\bibfnamefont {P.~R.}\ \bibnamefont {Brady}}, \bibinfo
  {author} {\bibfnamefont {D.~A.}\ \bibnamefont {Brown}}, \bibinfo {author}
  {\bibfnamefont {K.}~\bibnamefont {Cannon}}, \bibinfo {author} {\bibfnamefont
  {C.~D.}\ \bibnamefont {Capano}}, \bibinfo {author} {\bibfnamefont {J.~H.}\
  \bibnamefont {Clayton}}, \bibinfo {author} {\bibfnamefont {T.}~\bibnamefont
  {Cokelaer}}, \bibinfo {author} {\bibfnamefont {J.~D.~E.}\ \bibnamefont
  {Creighton}}, \bibinfo {author} {\bibfnamefont {T.}~\bibnamefont {Dent}},
  \bibinfo {author} {\bibfnamefont {A.}~\bibnamefont {Dietz}}, \bibinfo
  {author} {\bibfnamefont {S.}~\bibnamefont {Fairhurst}}, \bibinfo {author}
  {\bibfnamefont {N.}~\bibnamefont {Fotopoulos}}, \bibinfo {author}
  {\bibfnamefont {G.}~\bibnamefont {Gonz{\'{a} }lez}}, \bibinfo {author}
  {\bibfnamefont {C.}~\bibnamefont {Hanna}}, \bibinfo {author} {\bibfnamefont
  {I.~W.}\ \bibnamefont {Harry}}, \bibinfo {author} {\bibfnamefont
  {G.}~\bibnamefont {Jones}}, \bibinfo {author} {\bibfnamefont
  {D.}~\bibnamefont {Keppel}}, \bibinfo {author} {\bibfnamefont {D.~J.~A.}\
  \bibnamefont {McKechan}}, \bibinfo {author} {\bibfnamefont {L.}~\bibnamefont
  {Pekowsky}}, \bibinfo {author} {\bibfnamefont {S.}~\bibnamefont {Privitera}},
  \bibinfo {author} {\bibfnamefont {C.}~\bibnamefont {Robinson}}, \bibinfo
  {author} {\bibfnamefont {A.~C.}\ \bibnamefont {Rodriguez}}, \bibinfo {author}
  {\bibfnamefont {B.~S.}\ \bibnamefont {Sathyaprakash}}, \bibinfo {author}
  {\bibfnamefont {A.~S.}\ \bibnamefont {Sengupta}}, \bibinfo {author}
  {\bibfnamefont {M.}~\bibnamefont {Vallisneri}}, \bibinfo {author}
  {\bibfnamefont {R.}~\bibnamefont {Vaulin}},\ and\ \bibinfo {author}
  {\bibfnamefont {A.~J.}\ \bibnamefont {Weinstein}},\ }\bibfield  {title}
  {\bibinfo {title} {Searching for gravitational waves from binary
  coalescence},\ }\href {https://doi.org/10.1103%2Fphysrevd.87.024033}
  {\bibfield  {journal} {\bibinfo  {journal} {Physical Review D}\ }\textbf
  {\bibinfo {volume} {87}} (\bibinfo {year} {2013})}\BibitemShut {NoStop}%
\bibitem [{\citenamefont {Hesthaven}\ \emph {et~al.}(2015)\citenamefont
  {Hesthaven}, \citenamefont {Rozza},\ and\ \citenamefont {Stamm}}]{Jan-RB}%
  \BibitemOpen
  \bibfield  {author} {\bibinfo {author} {\bibfnamefont {J.~S.}\ \bibnamefont
  {Hesthaven}}, \bibinfo {author} {\bibfnamefont {G.}~\bibnamefont {Rozza}},\
  and\ \bibinfo {author} {\bibfnamefont {B.}~\bibnamefont {Stamm}},\ }\href
  {https://doi.org/10.1007/978-3-319-22470-1} {\emph {\bibinfo {title}
  {Certified Reduced Basis Methods for Parametrized Partial Differential
  Equations}}},\ \bibinfo {edition} {1st}\ ed.,\ Springer Briefs in
  Mathematics\ (\bibinfo  {publisher} {Springer},\ \bibinfo {address}
  {Switzerland},\ \bibinfo {year} {2015})\ p.\ \bibinfo {pages}
  {135}\BibitemShut {NoStop}%
\bibitem [{\citenamefont {Chen}\ \emph {et~al.}(2010)\citenamefont {Chen},
  \citenamefont {Hesthaven}, \citenamefont {Maday},\ and\ \citenamefont
  {Rodr\'{\i}guez}}]{Chen:2010:CRB:1958598.1958625}%
  \BibitemOpen
  \bibfield  {author} {\bibinfo {author} {\bibfnamefont {Y.}~\bibnamefont
  {Chen}}, \bibinfo {author} {\bibfnamefont {J.~S.}\ \bibnamefont {Hesthaven}},
  \bibinfo {author} {\bibfnamefont {Y.}~\bibnamefont {Maday}},\ and\ \bibinfo
  {author} {\bibfnamefont {J.}~\bibnamefont {Rodr\'{\i}guez}},\ }\bibfield
  {title} {\bibinfo {title} {Certified reduced basis methods and output bounds
  for the harmonic maxwell's equations},\ }\href
  {https://doi.org/10.1137/09075250X} {\bibfield  {journal} {\bibinfo
  {journal} {SIAM J. Sci. Comput.}\ }\textbf {\bibinfo {volume} {32}},\
  \bibinfo {pages} {970} (\bibinfo {year} {2010})}\BibitemShut {NoStop}%
\bibitem [{\citenamefont {Field}\ \emph {et~al.}(2011)\citenamefont {Field},
  \citenamefont {Galley}, \citenamefont {Herrmann}, \citenamefont {Hesthaven},
  \citenamefont {Ochsner},\ and\ \citenamefont {Tiglio}}]{Field:2011mf}%
  \BibitemOpen
  \bibfield  {author} {\bibinfo {author} {\bibfnamefont {S.~E.}\ \bibnamefont
  {Field}}, \bibinfo {author} {\bibfnamefont {C.~R.}\ \bibnamefont {Galley}},
  \bibinfo {author} {\bibfnamefont {F.}~\bibnamefont {Herrmann}}, \bibinfo
  {author} {\bibfnamefont {J.~S.}\ \bibnamefont {Hesthaven}}, \bibinfo {author}
  {\bibfnamefont {E.}~\bibnamefont {Ochsner}},\ and\ \bibinfo {author}
  {\bibfnamefont {M.}~\bibnamefont {Tiglio}},\ }\bibfield  {title} {\bibinfo
  {title} {{Reduced basis catalogs for gravitational wave templates}},\ }\href
  {https://doi.org/10.1103/PhysRevLett.106.221102} {\bibfield  {journal}
  {\bibinfo  {journal} {Phys. Rev. Lett.}\ }\textbf {\bibinfo {volume} {106}},\
  \bibinfo {pages} {221102} (\bibinfo {year} {2011})}\BibitemShut {NoStop}%
\bibitem [{\citenamefont {Prud'homme}\ \emph {et~al.}(2002)\citenamefont
  {Prud'homme}, \citenamefont {Rovas}, \citenamefont {Veroy}, \citenamefont
  {Machiels}, \citenamefont {Maday}, \citenamefont {Patera},\ and\
  \citenamefont {Turinici}}]{prud'homme:70}%
  \BibitemOpen
  \bibfield  {author} {\bibinfo {author} {\bibfnamefont {C.}~\bibnamefont
  {Prud'homme}}, \bibinfo {author} {\bibfnamefont {D.~V.}\ \bibnamefont
  {Rovas}}, \bibinfo {author} {\bibfnamefont {K.}~\bibnamefont {Veroy}},
  \bibinfo {author} {\bibfnamefont {L.}~\bibnamefont {Machiels}}, \bibinfo
  {author} {\bibfnamefont {Y.}~\bibnamefont {Maday}}, \bibinfo {author}
  {\bibfnamefont {A.~T.}\ \bibnamefont {Patera}},\ and\ \bibinfo {author}
  {\bibfnamefont {G.}~\bibnamefont {Turinici}},\ }\bibfield  {title} {\bibinfo
  {title} {Reliable real-time solution of parametrized partial differential
  equations: Reduced-basis output bound methods},\ }\href
  {https://doi.org/10.1115/1.1448332} {\bibfield  {journal} {\bibinfo
  {journal} {J. Fluids Eng.}\ }\textbf {\bibinfo {volume} {124}},\ \bibinfo
  {pages} {70} (\bibinfo {year} {2002})}\BibitemShut {NoStop}%
\bibitem [{\citenamefont {Quarteroni}\ \emph {et~al.}(2015)\citenamefont
  {Quarteroni}, \citenamefont {Manzoni},\ and\ \citenamefont
  {Negri}}]{quarteroni2015reduced}%
  \BibitemOpen
  \bibfield  {author} {\bibinfo {author} {\bibfnamefont {A.}~\bibnamefont
  {Quarteroni}}, \bibinfo {author} {\bibfnamefont {A.}~\bibnamefont
  {Manzoni}},\ and\ \bibinfo {author} {\bibfnamefont {F.}~\bibnamefont
  {Negri}},\ }\href {https://doi.org/10.1007/978-3-319-15431-2} {\emph
  {\bibinfo {title} {Reduced Basis Methods for Partial Differential Equations:
  An Introduction}}},\ UNITEXT\ (\bibinfo  {publisher} {Springer International
  Publishing},\ \bibinfo {year} {2015})\BibitemShut {NoStop}%
\bibitem [{\citenamefont {Varma}\ \emph
  {et~al.}(2019{\natexlab{a}})\citenamefont {Varma}, \citenamefont {Field},
  \citenamefont {Scheel}, \citenamefont {Blackman}, \citenamefont {Gerosa},
  \citenamefont {Stein}, \citenamefont {Kidder},\ and\ \citenamefont
  {Pfeiffer}}]{Varma:2019csw}%
  \BibitemOpen
  \bibfield  {author} {\bibinfo {author} {\bibfnamefont {V.}~\bibnamefont
  {Varma}}, \bibinfo {author} {\bibfnamefont {S.~E.}\ \bibnamefont {Field}},
  \bibinfo {author} {\bibfnamefont {M.~A.}\ \bibnamefont {Scheel}}, \bibinfo
  {author} {\bibfnamefont {J.}~\bibnamefont {Blackman}}, \bibinfo {author}
  {\bibfnamefont {D.}~\bibnamefont {Gerosa}}, \bibinfo {author} {\bibfnamefont
  {L.~C.}\ \bibnamefont {Stein}}, \bibinfo {author} {\bibfnamefont {L.~E.}\
  \bibnamefont {Kidder}},\ and\ \bibinfo {author} {\bibfnamefont {H.~P.}\
  \bibnamefont {Pfeiffer}},\ }\bibfield  {title} {\bibinfo {title} {{Surrogate
  models for precessing binary black hole simulations with unequal masses}},\
  }\href {https://doi.org/10.1103/PhysRevResearch.1.033015} {\bibfield
  {journal} {\bibinfo  {journal} {Phys. Rev. Research.}\ }\textbf {\bibinfo
  {volume} {1}},\ \bibinfo {pages} {033015} (\bibinfo {year}
  {2019}{\natexlab{a}})}\BibitemShut {NoStop}%
\bibitem [{GWSurrogate()}]{gwsurr}%
  \BibitemOpen
  GWSurrogate,\ \href {https://pypi.org/project/gwsurrogate/} {\bibinfo {title}
  {{GWSurrogate}}} (\bibinfo {year} {2014--2020}),\ \bibinfo {note} {accessed
  31 May 2021}\BibitemShut {NoStop}%
\bibitem [{\citenamefont {Field}\ \emph {et~al.}(2019)\citenamefont {Field},
  \citenamefont {Galley}, \citenamefont {Hesthaven}, \citenamefont {Kaye},
  \citenamefont {Tiglio}, \citenamefont {Blackman}, \citenamefont {Szil\'agyi},
  \citenamefont {Scheel}, \citenamefont {Hemberger}, \citenamefont {Schmidt},
  \citenamefont {Smith}, \citenamefont {Ott}, \citenamefont {Boyle},
  \citenamefont {Kidder}, \citenamefont {Pfeiffer},\ and\ \citenamefont
  {Varma}}]{catalog}%
  \BibitemOpen
  \bibfield  {author} {\bibinfo {author} {\bibfnamefont {S.~E.}\ \bibnamefont
  {Field}}, \bibinfo {author} {\bibfnamefont {C.~R.}\ \bibnamefont {Galley}},
  \bibinfo {author} {\bibfnamefont {J.~S.}\ \bibnamefont {Hesthaven}}, \bibinfo
  {author} {\bibfnamefont {J.}~\bibnamefont {Kaye}}, \bibinfo {author}
  {\bibfnamefont {M.}~\bibnamefont {Tiglio}}, \bibinfo {author} {\bibfnamefont
  {J.}~\bibnamefont {Blackman}}, \bibinfo {author} {\bibfnamefont
  {B.}~\bibnamefont {Szil\'agyi}}, \bibinfo {author} {\bibfnamefont {M.~A.}\
  \bibnamefont {Scheel}}, \bibinfo {author} {\bibfnamefont {D.~A.}\
  \bibnamefont {Hemberger}}, \bibinfo {author} {\bibfnamefont {P.}~\bibnamefont
  {Schmidt}}, \bibinfo {author} {\bibfnamefont {R.}~\bibnamefont {Smith}},
  \bibinfo {author} {\bibfnamefont {C.~D.}\ \bibnamefont {Ott}}, \bibinfo
  {author} {\bibfnamefont {M.}~\bibnamefont {Boyle}}, \bibinfo {author}
  {\bibfnamefont {L.~E.}\ \bibnamefont {Kidder}}, \bibinfo {author}
  {\bibfnamefont {H.~P.}\ \bibnamefont {Pfeiffer}},\ and\ \bibinfo {author}
  {\bibfnamefont {V.}~\bibnamefont {Varma}},\ }\href
  {https://doi.org/10.5281/zenodo.3629749} {\bibinfo {title} {Binary black-hole
  surrogate waveform catalog}} (\bibinfo {year} {2019})\BibitemShut {NoStop}%
\bibitem [{\citenamefont {Blackman}\ \emph
  {et~al.}(2017{\natexlab{a}})\citenamefont {Blackman}, \citenamefont {Field},
  \citenamefont {Scheel}, \citenamefont {Galley}, \citenamefont {Hemberger},
  \citenamefont {Schmidt},\ and\ \citenamefont {Smith}}]{PhysRevD.95.104023}%
  \BibitemOpen
  \bibfield  {author} {\bibinfo {author} {\bibfnamefont {J.}~\bibnamefont
  {Blackman}}, \bibinfo {author} {\bibfnamefont {S.~E.}\ \bibnamefont {Field}},
  \bibinfo {author} {\bibfnamefont {M.~A.}\ \bibnamefont {Scheel}}, \bibinfo
  {author} {\bibfnamefont {C.~R.}\ \bibnamefont {Galley}}, \bibinfo {author}
  {\bibfnamefont {D.~A.}\ \bibnamefont {Hemberger}}, \bibinfo {author}
  {\bibfnamefont {P.}~\bibnamefont {Schmidt}},\ and\ \bibinfo {author}
  {\bibfnamefont {R.}~\bibnamefont {Smith}},\ }\bibfield  {title} {\bibinfo
  {title} {A surrogate model of gravitational waveforms from numerical
  relativity simulations of precessing binary black hole mergers},\ }\href
  {https://doi.org/10.1103/PhysRevD.95.104023} {\bibfield  {journal} {\bibinfo
  {journal} {Phys. Rev. D}\ }\textbf {\bibinfo {volume} {95}},\ \bibinfo
  {pages} {104023} (\bibinfo {year} {2017}{\natexlab{a}})}\BibitemShut
  {NoStop}%
\bibitem [{\citenamefont {Blackman}\ \emph {et~al.}(2015)\citenamefont
  {Blackman}, \citenamefont {Field}, \citenamefont {Galley}, \citenamefont
  {Szil\'agyi}, \citenamefont {Scheel}, \citenamefont {Tiglio},\ and\
  \citenamefont {Hemberger}}]{PhysRevLett.115.121102}%
  \BibitemOpen
  \bibfield  {author} {\bibinfo {author} {\bibfnamefont {J.}~\bibnamefont
  {Blackman}}, \bibinfo {author} {\bibfnamefont {S.~E.}\ \bibnamefont {Field}},
  \bibinfo {author} {\bibfnamefont {C.~R.}\ \bibnamefont {Galley}}, \bibinfo
  {author} {\bibfnamefont {B.}~\bibnamefont {Szil\'agyi}}, \bibinfo {author}
  {\bibfnamefont {M.~A.}\ \bibnamefont {Scheel}}, \bibinfo {author}
  {\bibfnamefont {M.}~\bibnamefont {Tiglio}},\ and\ \bibinfo {author}
  {\bibfnamefont {D.~A.}\ \bibnamefont {Hemberger}},\ }\bibfield  {title}
  {\bibinfo {title} {Fast and accurate prediction of numerical relativity
  waveforms from binary black hole coalescences using surrogate models},\
  }\href {https://doi.org/10.1103/PhysRevLett.115.121102} {\bibfield  {journal}
  {\bibinfo  {journal} {Phys. Rev. Lett.}\ }\textbf {\bibinfo {volume} {115}},\
  \bibinfo {pages} {121102} (\bibinfo {year} {2015})}\BibitemShut {NoStop}%
\bibitem [{\citenamefont {Blackman}\ \emph
  {et~al.}(2017{\natexlab{b}})\citenamefont {Blackman}, \citenamefont {Field},
  \citenamefont {Scheel}, \citenamefont {Galley}, \citenamefont {Ott},
  \citenamefont {Boyle}, \citenamefont {Kidder}, \citenamefont {Pfeiffer},\
  and\ \citenamefont {Szil\'agyi}}]{PhysRevD.96.024058}%
  \BibitemOpen
  \bibfield  {author} {\bibinfo {author} {\bibfnamefont {J.}~\bibnamefont
  {Blackman}}, \bibinfo {author} {\bibfnamefont {S.~E.}\ \bibnamefont {Field}},
  \bibinfo {author} {\bibfnamefont {M.~A.}\ \bibnamefont {Scheel}}, \bibinfo
  {author} {\bibfnamefont {C.~R.}\ \bibnamefont {Galley}}, \bibinfo {author}
  {\bibfnamefont {C.~D.}\ \bibnamefont {Ott}}, \bibinfo {author} {\bibfnamefont
  {M.}~\bibnamefont {Boyle}}, \bibinfo {author} {\bibfnamefont {L.~E.}\
  \bibnamefont {Kidder}}, \bibinfo {author} {\bibfnamefont {H.~P.}\
  \bibnamefont {Pfeiffer}},\ and\ \bibinfo {author} {\bibfnamefont
  {B.}~\bibnamefont {Szil\'agyi}},\ }\bibfield  {title} {\bibinfo {title}
  {Numerical relativity waveform surrogate model for generically precessing
  binary black hole mergers},\ }\href
  {https://doi.org/10.1103/PhysRevD.96.024058} {\bibfield  {journal} {\bibinfo
  {journal} {Phys. Rev. D}\ }\textbf {\bibinfo {volume} {96}},\ \bibinfo
  {pages} {024058} (\bibinfo {year} {2017}{\natexlab{b}})}\BibitemShut
  {NoStop}%
\bibitem [{\citenamefont {Field}\ \emph {et~al.}(2014)\citenamefont {Field},
  \citenamefont {Galley}, \citenamefont {Hesthaven}, \citenamefont {Kaye},\
  and\ \citenamefont {Tiglio}}]{PhysRevX.4.031006}%
  \BibitemOpen
  \bibfield  {author} {\bibinfo {author} {\bibfnamefont {S.~E.}\ \bibnamefont
  {Field}}, \bibinfo {author} {\bibfnamefont {C.~R.}\ \bibnamefont {Galley}},
  \bibinfo {author} {\bibfnamefont {J.~S.}\ \bibnamefont {Hesthaven}}, \bibinfo
  {author} {\bibfnamefont {J.}~\bibnamefont {Kaye}},\ and\ \bibinfo {author}
  {\bibfnamefont {M.}~\bibnamefont {Tiglio}},\ }\bibfield  {title} {\bibinfo
  {title} {Fast prediction and evaluation of gravitational waveforms using
  surrogate models},\ }\href {https://doi.org/10.1103/PhysRevX.4.031006}
  {\bibfield  {journal} {\bibinfo  {journal} {Phys. Rev. X}\ }\textbf {\bibinfo
  {volume} {4}},\ \bibinfo {pages} {031006} (\bibinfo {year}
  {2014})}\BibitemShut {NoStop}%
\bibitem [{\citenamefont {Tiglio}\ and\ \citenamefont
  {Villanueva}(2022)}]{tiglio2021reduced}%
  \BibitemOpen
  \bibfield  {author} {\bibinfo {author} {\bibfnamefont {M.}~\bibnamefont
  {Tiglio}}\ and\ \bibinfo {author} {\bibfnamefont {A.}~\bibnamefont
  {Villanueva}},\ }\bibfield  {title} {\bibinfo {title} {Reduced order and
  surrogate models for gravitational waves},\ }\href
  {https://www.researchgate.net/publication/360205358_Reduced_order_and_surrogate_models_for_gravitational_waves}
  {\bibfield  {journal} {\bibinfo  {journal} {Living Reviews in Relativity}\
  }\textbf {\bibinfo {volume} {25}} (\bibinfo {year} {2022})}\BibitemShut
  {NoStop}%
\bibitem [{\citenamefont {Eftang}\ \emph {et~al.}(2010)\citenamefont {Eftang},
  \citenamefont {Patera},\ and\ \citenamefont {Ronquist}}]{Eftang:2010}%
  \BibitemOpen
  \bibfield  {author} {\bibinfo {author} {\bibfnamefont {J.~L.}\ \bibnamefont
  {Eftang}}, \bibinfo {author} {\bibfnamefont {A.~T.}\ \bibnamefont {Patera}},\
  and\ \bibinfo {author} {\bibfnamefont {E.~M.}\ \bibnamefont {Ronquist}},\
  }\bibfield  {title} {\bibinfo {title} {An hp certified reduced basis method
  for parametrized elliptic partial differential equations},\ }\href
  {https://dspace.mit.edu/handle/1721.1/58468} {\bibfield  {journal} {\bibinfo
  {journal} {SIAM J. Sci. Comput.}\ }\textbf {\bibinfo {volume} {32}},\
  \bibinfo {pages} {3170} (\bibinfo {year} {2010})}\BibitemShut {NoStop}%
\bibitem [{\citenamefont {Sarbach}\ and\ \citenamefont
  {Tiglio}(2012)}]{Sarbach2012}%
  \BibitemOpen
  \bibfield  {author} {\bibinfo {author} {\bibfnamefont {O.}~\bibnamefont
  {Sarbach}}\ and\ \bibinfo {author} {\bibfnamefont {M.}~\bibnamefont
  {Tiglio}},\ }\bibfield  {title} {\bibinfo {title} {Continuum and discrete
  initial-boundary value problems and einstein's field equations},\ }\href
  {https://doi.org/10.12942/lrr-2012-9} {\bibfield  {journal} {\bibinfo
  {journal} {Living Rev. Relativ.}\ }\textbf {\bibinfo {volume} {15}},\
  \bibinfo {eid} {9} (\bibinfo {year} {2012})}\BibitemShut {NoStop}%
\bibitem [{\citenamefont {Hesthaven}\ \emph {et~al.}(2007)\citenamefont
  {Hesthaven}, \citenamefont {Gottlieb},\ and\ \citenamefont
  {Gottlieb}}]{hesthaven2007spectral}%
  \BibitemOpen
  \bibfield  {author} {\bibinfo {author} {\bibfnamefont {J.}~\bibnamefont
  {Hesthaven}}, \bibinfo {author} {\bibfnamefont {S.}~\bibnamefont
  {Gottlieb}},\ and\ \bibinfo {author} {\bibfnamefont {D.}~\bibnamefont
  {Gottlieb}},\ }\href {https://doi.org/10.1017/CBO9780511618352.001} {\emph
  {\bibinfo {title} {Spectral Methods for Time-Dependent Problems}}},\
  Cambridge Monographs on Applied and Computational Mathematics\ (\bibinfo
  {publisher} {Cambridge University Press},\ \bibinfo {year}
  {2007})\BibitemShut {NoStop}%
\bibitem [{\citenamefont {Pinkus}(1985)}]{Pinkus}%
  \BibitemOpen
  \bibfield  {author} {\bibinfo {author} {\bibfnamefont {A.}~\bibnamefont
  {Pinkus}},\ }\href {https://doi.org/10.1007/978-3-642-69894-1} {\emph
  {\bibinfo {title} {N-widths in approximation theory}}}\ (\bibinfo
  {publisher} {Springer},\ \bibinfo {address} {Amsterdam},\ \bibinfo {year}
  {1985})\BibitemShut {NoStop}%
\bibitem [{\citenamefont {Magaril-Il'yaev}\ \emph {et~al.}(2001)\citenamefont
  {Magaril-Il'yaev}, \citenamefont {Osipenko},\ and\ \citenamefont
  {Tikhomirov}}]{MAGARILILYAEV200197}%
  \BibitemOpen
  \bibfield  {author} {\bibinfo {author} {\bibfnamefont {G.~G.}\ \bibnamefont
  {Magaril-Il'yaev}}, \bibinfo {author} {\bibfnamefont {K.~Y.}\ \bibnamefont
  {Osipenko}},\ and\ \bibinfo {author} {\bibfnamefont {V.~M.}\ \bibnamefont
  {Tikhomirov}},\ }\bibfield  {title} {\bibinfo {title} {On exact values of
  n-widths in a hilbert space},\ }\href
  {https://doi.org/10.1006/jath.2000.3497} {\bibfield  {journal} {\bibinfo
  {journal} {J. Approxim. Theory}\ }\textbf {\bibinfo {volume} {108}},\
  \bibinfo {pages} {97} (\bibinfo {year} {2001})}\BibitemShut {NoStop}%
\bibitem [{\citenamefont {{Field}}\ \emph {et~al.}(2012)\citenamefont
  {{Field}}, \citenamefont {{Galley}},\ and\ \citenamefont
  {{Ochsner}}}]{Beat2012}%
  \BibitemOpen
  \bibfield  {author} {\bibinfo {author} {\bibfnamefont {S.~E.}\ \bibnamefont
  {{Field}}}, \bibinfo {author} {\bibfnamefont {C.~R.}\ \bibnamefont
  {{Galley}}},\ and\ \bibinfo {author} {\bibfnamefont {E.}~\bibnamefont
  {{Ochsner}}},\ }\bibfield  {title} {\bibinfo {title} {{Towards beating the
  curse of dimensionality for gravitational waves using reduced basis}},\
  }\href {https://inspirehep.net/literature/1116236} {\bibfield  {journal}
  {\bibinfo  {journal} {\prd}\ }\textbf {\bibinfo {volume} {86}} (\bibinfo
  {year} {2012})}\BibitemShut {NoStop}%
\bibitem [{\citenamefont {DeVore}\ \emph {et~al.}(2013)\citenamefont {DeVore},
  \citenamefont {Petrova},\ and\ \citenamefont {Wojtaszczyk}}]{Devore2012}%
  \BibitemOpen
  \bibfield  {author} {\bibinfo {author} {\bibfnamefont {R.}~\bibnamefont
  {DeVore}}, \bibinfo {author} {\bibfnamefont {G.}~\bibnamefont {Petrova}},\
  and\ \bibinfo {author} {\bibfnamefont {P.}~\bibnamefont {Wojtaszczyk}},\
  }\bibfield  {title} {\bibinfo {title} {Greedy algorithms for reduced bases in
  banach spaces},\ }\href {https://doi.org/10.1007/s00365-013-9186-2}
  {\bibfield  {journal} {\bibinfo  {journal} {Constructive Approximation}\
  }\textbf {\bibinfo {volume} {37}},\ \bibinfo {pages} {455} (\bibinfo {year}
  {2013})}\BibitemShut {NoStop}%
\bibitem [{\citenamefont {Buffa}\ \emph {et~al.}(2012)\citenamefont {Buffa},
  \citenamefont {Maday}, \citenamefont {Patera}, \citenamefont {Prud'homme},\
  and\ \citenamefont {Turinici}}]{Buffa2012}%
  \BibitemOpen
  \bibfield  {author} {\bibinfo {author} {\bibfnamefont {A.}~\bibnamefont
  {Buffa}}, \bibinfo {author} {\bibfnamefont {Y.}~\bibnamefont {Maday}},
  \bibinfo {author} {\bibfnamefont {A.}~\bibnamefont {Patera}}, \bibinfo
  {author} {\bibfnamefont {C.}~\bibnamefont {Prud'homme}},\ and\ \bibinfo
  {author} {\bibfnamefont {G.}~\bibnamefont {Turinici}},\ }\bibfield  {title}
  {\bibinfo {title} {A priori convergence of the greedy algorithm for the
  parametrized reduced basis method},\ }\href
  {https://doi.org/10.1051/m2an/2011056} {\bibfield  {journal} {\bibinfo
  {journal} {ESAIM: Mathematical Modelling and Numerical Analysis}\ }\textbf
  {\bibinfo {volume} {46}} (\bibinfo {year} {2012})}\BibitemShut {NoStop}%
\bibitem [{\citenamefont {Binev}\ \emph {et~al.}(2011)\citenamefont {Binev},
  \citenamefont {Cohen}, \citenamefont {Dahmen}, \citenamefont {DeVore},
  \citenamefont {Petrova},\ and\ \citenamefont
  {Wojtaszczyk}}]{Binev10convergencerates}%
  \BibitemOpen
  \bibfield  {author} {\bibinfo {author} {\bibfnamefont {P.}~\bibnamefont
  {Binev}}, \bibinfo {author} {\bibfnamefont {A.}~\bibnamefont {Cohen}},
  \bibinfo {author} {\bibfnamefont {W.}~\bibnamefont {Dahmen}}, \bibinfo
  {author} {\bibfnamefont {R.~A.}\ \bibnamefont {DeVore}}, \bibinfo {author}
  {\bibfnamefont {G.}~\bibnamefont {Petrova}},\ and\ \bibinfo {author}
  {\bibfnamefont {P.}~\bibnamefont {Wojtaszczyk}},\ }\bibfield  {title}
  {\bibinfo {title} {Convergence rates for greedy algorithms in reduced basis
  methods},\ }\href
  {https://www.researchgate.net/publication/220132479_Convergence_Rates_for_Greedy_Algorithms_in_Reduced_Basis_Methods}
  {\bibfield  {journal} {\bibinfo  {journal} {SIAM J. Math. Analysis}\ }\textbf
  {\bibinfo {volume} {43}},\ \bibinfo {pages} {1457} (\bibinfo {year}
  {2011})}\BibitemShut {NoStop}%
\bibitem [{\citenamefont {Caudill}\ \emph {et~al.}(2012)\citenamefont
  {Caudill}, \citenamefont {Field}, \citenamefont {Galley}, \citenamefont
  {Herrmann},\ and\ \citenamefont {Tiglio}}]{Caudill:2011kv}%
  \BibitemOpen
  \bibfield  {author} {\bibinfo {author} {\bibfnamefont {S.}~\bibnamefont
  {Caudill}}, \bibinfo {author} {\bibfnamefont {S.~E.}\ \bibnamefont {Field}},
  \bibinfo {author} {\bibfnamefont {C.~R.}\ \bibnamefont {Galley}}, \bibinfo
  {author} {\bibfnamefont {F.}~\bibnamefont {Herrmann}},\ and\ \bibinfo
  {author} {\bibfnamefont {M.}~\bibnamefont {Tiglio}},\ }\bibfield  {title}
  {\bibinfo {title} {{Reduced Basis representations of multi-mode black hole
  ringdown gravitational waves}},\ }\href
  {https://iopscience.iop.org/article/10.1088/0264-9381/29/9/095016} {\bibfield
   {journal} {\bibinfo  {journal} {Class. Quant. Grav.}\ }\textbf {\bibinfo
  {volume} {29}},\ \bibinfo {pages} {095016} (\bibinfo {year}
  {2012})}\BibitemShut {NoStop}%
\bibitem [{\citenamefont {Taylor}(1978)}]{taylor_1978}%
  \BibitemOpen
  \bibfield  {author} {\bibinfo {author} {\bibfnamefont {J.~M.}\ \bibnamefont
  {Taylor}},\ }\bibfield  {title} {\bibinfo {title} {The condition of gram
  matrices and related problems},\ }\href
  {https://doi.org/10.1017/S030821050001012X} {\bibfield  {journal} {\bibinfo
  {journal} {Proc. R. Soc. Edinburgh: Sect. A Math.}\ }\textbf {\bibinfo
  {volume} {80}},\ \bibinfo {pages} {45–56} (\bibinfo {year}
  {1978})}\BibitemShut {NoStop}%
\bibitem [{\citenamefont {Hoffmann}(1989)}]{Hoffmann_IMGS}%
  \BibitemOpen
  \bibfield  {author} {\bibinfo {author} {\bibfnamefont {W.}~\bibnamefont
  {Hoffmann}},\ }\bibfield  {title} {\bibinfo {title} {Iterative algorithms for
  gram-schmidt orthogonalization},\ }\href {https://doi.org/10.1007/BF02241222}
  {\bibfield  {journal} {\bibinfo  {journal} {Computing}\ }\textbf {\bibinfo
  {volume} {41}},\ \bibinfo {pages} {335} (\bibinfo {year} {1989})}\BibitemShut
  {NoStop}%
\bibitem [{\citenamefont {Villanueva}\ \emph {et~al.}(2021)\citenamefont
  {Villanueva}, \citenamefont {Beroiz}, \citenamefont {Cabral}, \citenamefont
  {Chalela},\ and\ \citenamefont {Dominguez}}]{arby}%
  \BibitemOpen
  \bibfield  {author} {\bibinfo {author} {\bibfnamefont {A.}~\bibnamefont
  {Villanueva}}, \bibinfo {author} {\bibfnamefont {M.}~\bibnamefont {Beroiz}},
  \bibinfo {author} {\bibfnamefont {J.}~\bibnamefont {Cabral}}, \bibinfo
  {author} {\bibfnamefont {M.}~\bibnamefont {Chalela}},\ and\ \bibinfo {author}
  {\bibfnamefont {M.}~\bibnamefont {Dominguez}},\ }\href
  {https://arxiv.org/abs/2108.01305} {\bibinfo {title} {Arby - fast data-driven
  surrogates}} (\bibinfo {year} {2021})\BibitemShut {NoStop}%
\bibitem [{\citenamefont {James}\ \emph {et~al.}(2013)\citenamefont {James},
  \citenamefont {Witten}, \citenamefont {Hastie},\ and\ \citenamefont
  {Tibshirani}}]{James2013}%
  \BibitemOpen
  \bibfield  {author} {\bibinfo {author} {\bibfnamefont {G.}~\bibnamefont
  {James}}, \bibinfo {author} {\bibfnamefont {D.}~\bibnamefont {Witten}},
  \bibinfo {author} {\bibfnamefont {T.}~\bibnamefont {Hastie}},\ and\ \bibinfo
  {author} {\bibfnamefont {R.}~\bibnamefont {Tibshirani}},\ }\href
  {https://www.statlearning.com/} {\emph {\bibinfo {title} {An Introduction to
  Statistical Learning: with Applications in R}}}\ (\bibinfo  {publisher}
  {Springer},\ \bibinfo {year} {2013})\BibitemShut {NoStop}%
\bibitem [{\citenamefont {Varma}\ \emph
  {et~al.}(2019{\natexlab{b}})\citenamefont {Varma}, \citenamefont {Field},
  \citenamefont {Scheel}, \citenamefont {Blackman}, \citenamefont {Kidder},\
  and\ \citenamefont {Pfeiffer}}]{PhysRevD.99.064045}%
  \BibitemOpen
  \bibfield  {author} {\bibinfo {author} {\bibfnamefont {V.}~\bibnamefont
  {Varma}}, \bibinfo {author} {\bibfnamefont {S.~E.}\ \bibnamefont {Field}},
  \bibinfo {author} {\bibfnamefont {M.~A.}\ \bibnamefont {Scheel}}, \bibinfo
  {author} {\bibfnamefont {J.}~\bibnamefont {Blackman}}, \bibinfo {author}
  {\bibfnamefont {L.~E.}\ \bibnamefont {Kidder}},\ and\ \bibinfo {author}
  {\bibfnamefont {H.~P.}\ \bibnamefont {Pfeiffer}},\ }\bibfield  {title}
  {\bibinfo {title} {Surrogate model of hybridized numerical relativity binary
  black hole waveforms},\ }\href {https://doi.org/10.1103/PhysRevD.99.064045}
  {\bibfield  {journal} {\bibinfo  {journal} {Phys. Rev. D}\ }\textbf {\bibinfo
  {volume} {99}},\ \bibinfo {pages} {064045} (\bibinfo {year}
  {2019}{\natexlab{b}})}\BibitemShut {NoStop}%
\bibitem [{\citenamefont {Field}\ \emph {et~al.}(2012)\citenamefont {Field},
  \citenamefont {Galley},\ and\ \citenamefont {Ochsner}}]{Field:2012if}%
  \BibitemOpen
  \bibfield  {author} {\bibinfo {author} {\bibfnamefont {S.~E.}\ \bibnamefont
  {Field}}, \bibinfo {author} {\bibfnamefont {C.~R.}\ \bibnamefont {Galley}},\
  and\ \bibinfo {author} {\bibfnamefont {E.}~\bibnamefont {Ochsner}},\
  }\bibfield  {title} {\bibinfo {title} {{Towards beating the curse of
  dimensionality for gravitational waves using Reduced Basis}},\ }\href
  {https://inspirehep.net/literature/1116236} {\bibfield  {journal} {\bibinfo
  {journal} {Phys. Rev.}\ }\textbf {\bibinfo {volume} {D86}},\ \bibinfo {pages}
  {084046} (\bibinfo {year} {2012})}\BibitemShut {NoStop}%
\bibitem [{\citenamefont {Antil}\ \emph {et~al.}(2013)\citenamefont {Antil},
  \citenamefont {Field}, \citenamefont {Herrmann}, \citenamefont {Nochetto},\
  and\ \citenamefont {Tiglio}}]{antil2012two}%
  \BibitemOpen
  \bibfield  {author} {\bibinfo {author} {\bibfnamefont {H.}~\bibnamefont
  {Antil}}, \bibinfo {author} {\bibfnamefont {S.~E.}\ \bibnamefont {Field}},
  \bibinfo {author} {\bibfnamefont {F.}~\bibnamefont {Herrmann}}, \bibinfo
  {author} {\bibfnamefont {R.~H.}\ \bibnamefont {Nochetto}},\ and\ \bibinfo
  {author} {\bibfnamefont {M.}~\bibnamefont {Tiglio}},\ }\bibfield  {title}
  {\bibinfo {title} {Two-step greedy algorithm for reduced order quadratures},\
  }\href {https://doi.org/10.1007/s10915-013-9722-z} {\bibfield  {journal}
  {\bibinfo  {journal} {J. Sci. Comput.}\ }\textbf {\bibinfo {volume} {57}},\
  \bibinfo {pages} {604} (\bibinfo {year} {2013})}\BibitemShut {NoStop}%
\bibitem [{\citenamefont {Morisaki}\ and\ \citenamefont
  {Raymond}(2020)}]{Morisaki_2020}%
  \BibitemOpen
  \bibfield  {author} {\bibinfo {author} {\bibfnamefont {S.}~\bibnamefont
  {Morisaki}}\ and\ \bibinfo {author} {\bibfnamefont {V.}~\bibnamefont
  {Raymond}},\ }\bibfield  {title} {\bibinfo {title} {{Rapid Parameter
  Estimation of Gravitational Waves from Binary Neutron Star Coalescence using
  Focused Reduced Order Quadrature}},\ }\href
  {https://doi.org/10.1103/PhysRevD.102.104020} {\bibfield  {journal} {\bibinfo
   {journal} {Phys. Rev. D}\ }\textbf {\bibinfo {volume} {102}},\ \bibinfo
  {pages} {104020} (\bibinfo {year} {2020})}\BibitemShut {NoStop}%
\bibitem [{\citenamefont {Canizares}\ \emph {et~al.}(2015)\citenamefont
  {Canizares}, \citenamefont {Field}, \citenamefont {Gair}, \citenamefont
  {Raymond}, \citenamefont {Smith},\ and\ \citenamefont
  {Tiglio}}]{PhysRevLett.114.071104}%
  \BibitemOpen
  \bibfield  {author} {\bibinfo {author} {\bibfnamefont {P.}~\bibnamefont
  {Canizares}}, \bibinfo {author} {\bibfnamefont {S.~E.}\ \bibnamefont
  {Field}}, \bibinfo {author} {\bibfnamefont {J.}~\bibnamefont {Gair}},
  \bibinfo {author} {\bibfnamefont {V.}~\bibnamefont {Raymond}}, \bibinfo
  {author} {\bibfnamefont {R.}~\bibnamefont {Smith}},\ and\ \bibinfo {author}
  {\bibfnamefont {M.}~\bibnamefont {Tiglio}},\ }\bibfield  {title} {\bibinfo
  {title} {Accelerated gravitational wave parameter estimation with reduced
  order modeling},\ }\href {https://doi.org/10.1103/PhysRevLett.114.071104}
  {\bibfield  {journal} {\bibinfo  {journal} {Phys. Rev. Lett.}\ }\textbf
  {\bibinfo {volume} {114}},\ \bibinfo {pages} {071104} (\bibinfo {year}
  {2015})}\BibitemShut {NoStop}%
\bibitem [{\citenamefont {Smith}\ \emph {et~al.}(2016)\citenamefont {Smith},
  \citenamefont {Field}, \citenamefont {Blackburn}, \citenamefont {Haster},
  \citenamefont {P\"urrer}, \citenamefont {Raymond},\ and\ \citenamefont
  {Schmidt}}]{PhysRevD.94.044031}%
  \BibitemOpen
  \bibfield  {author} {\bibinfo {author} {\bibfnamefont {R.}~\bibnamefont
  {Smith}}, \bibinfo {author} {\bibfnamefont {S.~E.}\ \bibnamefont {Field}},
  \bibinfo {author} {\bibfnamefont {K.}~\bibnamefont {Blackburn}}, \bibinfo
  {author} {\bibfnamefont {C.-J.}\ \bibnamefont {Haster}}, \bibinfo {author}
  {\bibfnamefont {M.}~\bibnamefont {P\"urrer}}, \bibinfo {author}
  {\bibfnamefont {V.}~\bibnamefont {Raymond}},\ and\ \bibinfo {author}
  {\bibfnamefont {P.}~\bibnamefont {Schmidt}},\ }\bibfield  {title} {\bibinfo
  {title} {Fast and accurate inference on gravitational waves from precessing
  compact binaries},\ }\href {https://doi.org/10.1103/PhysRevD.94.044031}
  {\bibfield  {journal} {\bibinfo  {journal} {Phys. Rev. D}\ }\textbf {\bibinfo
  {volume} {94}},\ \bibinfo {pages} {044031} (\bibinfo {year}
  {2016})}\BibitemShut {NoStop}%
\bibitem [{\citenamefont {Pretorius}\ and\ \citenamefont
  {Khurana}(2007)}]{Pretorius_2007}%
  \BibitemOpen
  \bibfield  {author} {\bibinfo {author} {\bibfnamefont {F.}~\bibnamefont
  {Pretorius}}\ and\ \bibinfo {author} {\bibfnamefont {D.}~\bibnamefont
  {Khurana}},\ }\bibfield  {title} {\bibinfo {title} {Black hole mergers and
  unstable circular orbits},\ }\href
  {https://doi.org/10.1088/0264-9381/24/12/s07} {\bibfield  {journal} {\bibinfo
   {journal} {Classical and Quantum Gravity}\ }\textbf {\bibinfo {volume}
  {24}},\ \bibinfo {pages} {S83} (\bibinfo {year} {2007})}\BibitemShut
  {NoStop}%
\bibitem [{\citenamefont {Shibata}(2015)}]{Shibata_NR}%
  \BibitemOpen
  \bibfield  {author} {\bibinfo {author} {\bibfnamefont {M.}~\bibnamefont
  {Shibata}},\ }\href {https://doi.org/10.1142/9692} {\emph {\bibinfo {title}
  {Numerical Relativity}}},\ 100 Years of General Relativity\ (\bibinfo
  {publisher} {World Scientific Publishing Company},\ \bibinfo {year}
  {2015})\BibitemShut {NoStop}%
\end{thebibliography}%

\end{document}